\RequirePackage{lineno}
\documentclass[12pt,onecolumn]{article}	
\usepackage[top=1in, bottom=1in, left=1in, right=1in]{geometry}
\usepackage{graphicx}
\usepackage{amssymb}
\usepackage{amsmath}  				
\usepackage{graphicx,float,wrapfig} 	
\usepackage{color}					
\usepackage[colorlinks,linktocpage=true]{hyperref}		
\usepackage[font=small,labelfont=bf,textfont=sf,
labelsep=quad]{caption}				
\usepackage{titlesec}

\titleformat{\section}{\normalfont\fontsize{14}{15}\bfseries}{\thesection}{1em}{}
\titleformat{\subsection}{\normalfont\fontsize{12}{15}\bfseries}{\thesubsection}{1em}{}
\titleformat{\subsubsection}{\normalfont\fontsize{11}{15}\bfseries}{\thesubsubsection}{1em}{}

\usepackage{gensymb}
\usepackage{caption}
\usepackage{subfig}

\hypersetup{						
  citecolor= black,
  linkcolor=blue,
}

\usepackage{setspace}   

\usepackage [english]{babel}
\usepackage [english = american]{csquotes}
\MakeOuterQuote{"}

 \usepackage[protrusion=false,expansion=false]{microtype}
  
 \usepackage[colorinlistoftodos]{todonotes}
\usepackage{soul}

\let\baraccent=\= 
\renewcommand{\=}[1]{\stackrel{#1}{=}} 

\begin{document}
\begin{center}
\Large{\textbf{\textit{Schistosoma mansoni} cercariae exploit an elastohydrodynamic coupling to swim efficiently}}\\
\end{center}
\noindent
\begin{center}
Deepak Krishnamurthy$^1$, Georgios Katsikis$^1$, Arjun Bhargava$^2$ \& Manu Prakash$^{3\ast}$\\
\small{$^{1}$Department of Mechanical Engineering, $^{2}$Department of Applied Physics} \\
\small{$^{3}$Department of Bioengineering,} \\
\small{Stanford University, Stanford, CA}\\
\small{$^\ast$To whom correspondence should be addressed; E-mail:  manup@stanford.edu}
\end{center}

\noindent\textbf{Keywords}\\
Motility, Parasites, Swimming, Hydrodynamics, Low Reynolds number, Schistosomiasis \\



\begin{spacing}{1.5}
\noindent\textbf{Abstract}\\
\footnotesize
The motility of many parasites is critical for the infection process of their host, as exemplified by the transmission cycle of the blood fluke \textit{Schistosoma mansoni} \cite{Haas1992}. In their human infectious stage, immature, submillimetre-scale forms of the parasite known as cercariae swim in freshwater and infect human hosts by penetrating through the skin \cite{Haas1992, Haas2003}. This infection causes Schistosomiasis, a parasitic disease that is comparable to malaria in terms of global socio-economic impact \cite{hotez2009neglected,Hotez2009}. Given that cercariae do not feed and hence have a finite lifetime of around 12 hours \cite{lawson1980survival,Whitfield2003}, efficient motility is crucial for the parasite's survival and transmission of the Schistosomiasis disease. However, a first-principles understanding of how cercariae swim is completely lacking. Via a combined experimental, theoretical and robotics based approach - we demonstrate that cercariae efficiently propel themselves against gravity by exploiting a unique elastohydrodynamic coupling. We show that cercariae beat their tail in a periodic fashion while maintaining a fixed flexibility near their posterior and anterior ends. The flexibility in these regions allows an interaction between the fluid drag and bending resistance --- an elastohydrodynamic coupling, to naturally break time-reversal symmetry and enable locomotion at small length-scales \cite{Purcell1977}. We present a theoretical model, a `T-swimmer', which captures the key swimming phenotype of cercariae. We further validate our results experimentally through a macro-scale robotic realization of the `T-swimmer', explaining the unique forked-tail geometry of cercariae. Finally, we find that cercariae maintain the flexibility at their posterior and anterior ends at an optimal regime for efficient swimming, as predicted by our theoretical model. We anticipate that our work sets the ground for linking the swimming of \textit{S. mansoni} cercariae to disease transmission and enables explorations of novel strategies for Schistosomiasis control and prevention.
\end{spacing}
\doublespacing
\normalsize
\newpage
\section*{Introduction}
\label{sec:intro}
In their natural habitat of rivers and ponds, \textit {S. mansoni} cercariae are released from intermediate snail hosts and aggregate below the water surface to increase their chances of finding a human host \cite{haas2008selection,combes1994behaviours} (Fig. \ref{fig:figure_1}a). The cercariae are around $500 ~\mu m$ long and consist of an anterior head, a slender tail and a posterior bifurcation of the tail called the `furca', henceforth termed the fork (Fig. \ref{fig:figure_1}b) \cite{Graefe1967,nuttman1975structure}. Being negatively buoyant, they need to swim against gravity in order to stay near the water surface, which they accomplish by using their fork \cite{Graefe1967,nuttman1975structure}. The fork is an uncommon appendage, not found in any well-studied swimming microorganisms such as bacteria \cite{berg_book}, algae or ciliates \cite{lauga_powers,ganesh_2012_review}. Also, unlike the aforementioned microorganisms, cercariae do not feed \cite{lawson1980survival,Whitfield2003}. Hence their swimming efficiency is an important determinant of their ability to survive longer to find and infect humans. For such an important human parasite, while a body of work has focused on the behavioral \cite{Haas1992,Haas2003} and qualitative aspects \cite{Graefe1967, bundy1981swimming} of cercarial motility, the swimming mechanics of cercariae remains unknown.  

\section*{Macroscale swimming behaviour}
\label{sec:bulk_swimming}
\begin{figure}
\begin{center}
\includegraphics[width=1.0\textwidth]{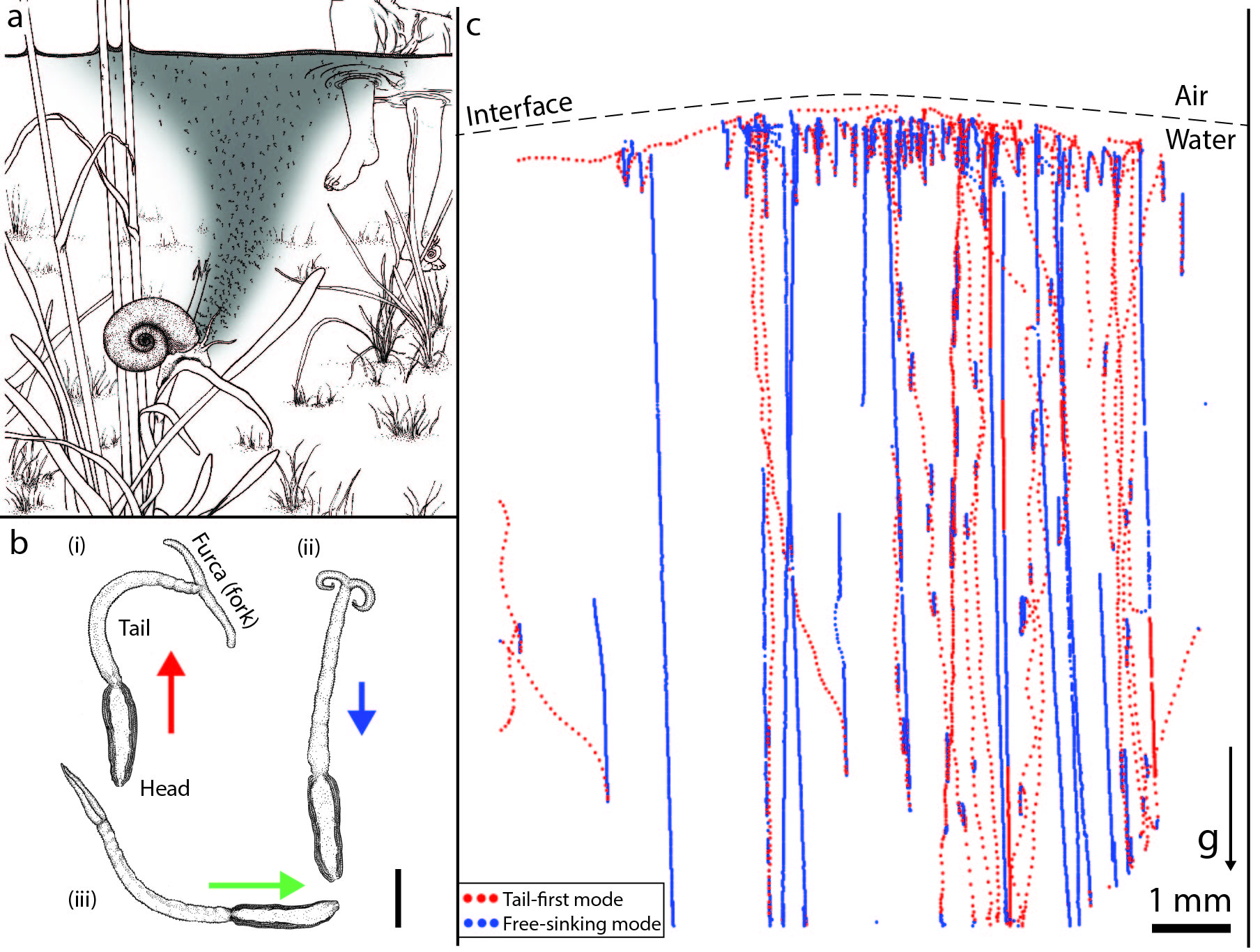} 
\caption{\label{fig:figure_1}\textbf{Macroscale swimming behaviour of \textit{Schistosoma mansoni} cercariae.} \textbf{a.} Artistic rendering (not-to-scale) of a typical transmission site where cercariae (immature, human infectious of the parasites) emerge into freshwater from snails and aggregate near the water surface \cite{haas2008selection}. \textbf{b,} Artistic rendering of the three distinct modes displayed by cercariae in their intermittent swimming behaviour termed (i) tail-first mode and (ii) free-sinking mode and (iii) head-first mode. Arrows indicate the direction of motion. Scale bar $100 ~ \mu m$. \textbf{c.} Snapshot of cercariae swimming in a chamber (Methods) showing tail-first swimming (red) and sinking (blue) trajectories for $10$ minutes. The resulting tracks of organisms have a columnar nature with very little drift in the horizontal directions. This is due to the fact that cercariae are bottom-heavy swimmers (Supplementary Fig. \ref{fig:SI_figure_reorient}) and undergo a gravitational realignment such that their longitudinal axis is aligned with the vertical with the head pointing down. Scale bar is $1 ~ mm$.
}
\end{center}
\end{figure}
To mimic the natural habitat of cercariae we conducted laboratory experiments with multiple freely swimming specimens in a $4 ~cm \times 1 ~cm \times 1.4 ~ mm$ chamber which is much larger than their own size ($\approx 500 ~ \mu m)$. We observed that cercariae display an intermittent swimming behaviour consisting of three modes \cite{Graefe1967,nuttman1975structure} (Fig. \ref{fig:figure_1} b): A tail-first mode in which they swim up against gravity with their fork fully extended (Fig. \ref{fig:figure_1} b (i)), a free-sinking mode on account of being negatively buoyant where the fork is partially extended (Fig. \ref{fig:figure_1}b (ii)) and a head-first mode which is used to penetrate through the skin, where the fork is folded back (Fig. \ref{fig:figure_1}b (iii)). We analysed trajectories ($n=2774$ segments; Methods) of individual cercariae and obtained the relative time durations spent in each mode. We found that cercariae spend a majority of their time in the free-sinking mode (83 \% of the total time) with a mean sinking speed $\langle V_{sinking} \rangle = 0.1 ~ mm ~ s^{-1}$ followed by the tail-first swimming mode (17 \%) with $\langle V_{tail-first} \rangle =0.7 ~ mm ~ s^{-1}$. The head-first mode is often triggered by chemical cues \cite{Haas2008,Brachs2008,Haas2002} which are absent in our current experiments, leading to very few observed trajectories ($< 0.1 \%$). Despite the fact that cercariae spend longer periods of time freely sinking, we expect that most of the energy consumption occurs during the tail-first mode \cite{lawson1980survival}, since this involves active swimming. Thus the swimming efficiency of the tail-first mode directly affects the ability of cercariae to aggregate near the water surface to increase chances of encountering a human host. Also, given that the swimming gait of cercariae in the tail-first mode is distinct from that of well-studied microorganisms \cite{ganesh_2012_review}, it is of fundamental interest to understand its mechanics.
\section*{Swimming kinematics: Tail-first swimming}
\label{sec:tail_swimming}
\begin{figure}
\begin{center}
\includegraphics[width=1.0\textwidth]{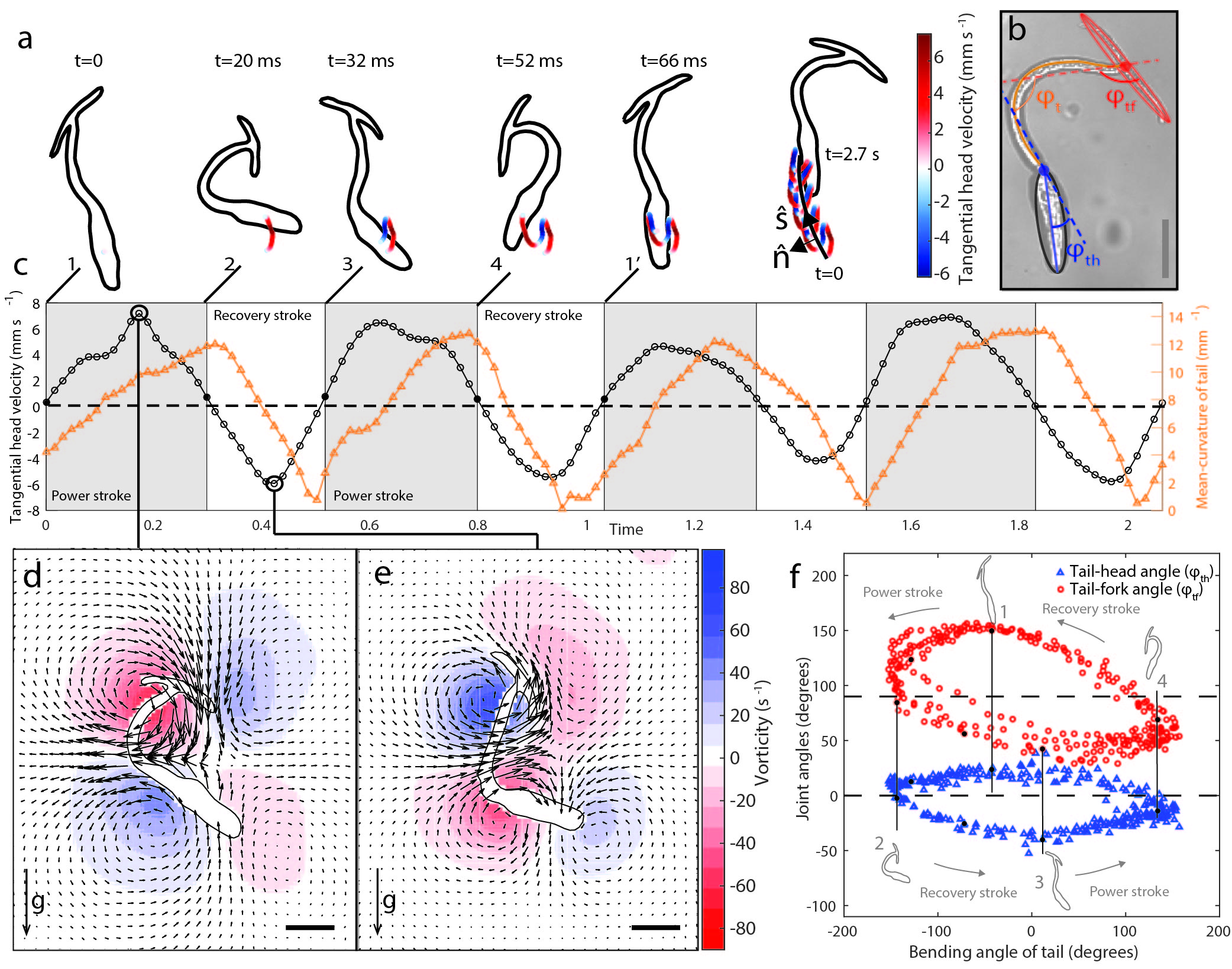}
\caption{\label{fig:figure_2} \textbf{Shape kinematics of \textit{S. mansoni} cercariae swimming in the tail-first mode.} \textbf{a,} Extracted dynamic shapes from high speed videos of cercariae at different time-points during a swimming cycle (Methods). The trajectory of the head is colour-coded based on the instantaneous velocity magnitude along the local tangent ($\boldsymbol{\hat{s}}$) to the mean swimming trajectory shown as a solid black line in the last panel. \textbf{b,} Illustration of kinematic parameters: the angle at the tail-head joint $\phi_{th}$ shown in blue,  the angle at the tail-fork joint $\phi_{tf}$ shown in red and an effective bending angle of the tail $\phi_t$ shown in orange. \textbf{c,} Time series plot of the tangential head velocity along $\boldsymbol{\hat{s}}$ (left axis) demonstrates forward motion although mean-curvature of the tail (right axis) is symmetric, seemingly appearing to be a reciprocal stroke. Gray regions ($1 \rightarrow 2, 3 \rightarrow 4$ etc) correspond to power-strokes where the head of the organism has a tangential velocity in the mean swimming direction, while white regions ($2 \rightarrow 3, 4 \rightarrow 1'$ etc) are recovery-strokes where the head moves in the opposite direction. Time is normalized by $1/f$, where $f = 15.6 ~ Hz$ is the beat frequency. \textbf{d, e} Snapshots of the transient flow velocity exhibiting reversing stagnation points. Vorticity fields obtained using particle-image-velocimetry (PIV) during the power-stroke and recovery-stroke, respectively. The regions of vorticity remain attached to the organism and are not shed. \textbf{f,} Phase plot of $\phi_{th}$ and $\phi_{tf}$ as a function of $\phi_t$ clearly shows a finite area hence revealing the role of tail-fork and tail-head joints in producing a non-reciprocal swimming stroke, required for low-Reynolds-number swimming. Scale bars are $100 ~ \mu m$.
}
\end{center}
\end{figure}
To study the tail-first swimming mode we conducted high-speed imaging experiments (Methods) on individual cercariae. We observed that cercariae swim using a periodic gait, wherein they beat their tail from side-to-side with a frequency $f$ in the range $15 - 20 Hz$ (Fig. \ref{fig:figure_2}a). The plane in which the tail beats varies much slower ($ \approx 250 ~ ms$) than a swimming cycle  ($t_{cycle}=1/f \approx 50 ~ ms$), hence justifying a two-dimensional description of the swimming motion.

For understanding the hydrodynamics of cercarial swimming, we estimated the relevant Reynolds number $Re = \langle V_{tail-first} \rangle L / \nu \approx 0.3$, where $L \approx 500 ~ \mu m$ is the organism length and $\nu = 10^{-6} ~ m^2 s^{-1}$ is the kinematic viscosity of the ambient fluid (water). This $Re < 1$ indicates the relative dominance of viscous over inertial forces in the swimming. To confirm this we measured the flow-field around the organism using Particle-Image-Velocimetry (PIV) (Methods) and observed regions of recirculating fluid (vortices) which remain adjacent to the organism (Fig. \ref{fig:figure_2}d and e). The flow field exhibits stagnation points that reverse orientation in each cycle. The vorticity (a measure of local fluid rotation rate \cite{lealbook}) shows peaks at the surface of the organism, with no vortex shedding: a characteristic feature of flow dominated by viscous forces \cite{sznitman2010propulsive}. Based on nature of the vorticity field and other flow metrics (Supplementary Fig. \ref{fig:SI_figure_1a}, \ref{fig:SI_figure_1b}, \ref{fig:SI_figure_1c}) we concluded that inertial forces are negligible in cercarial swimming. We therefore studied the hydrodynamics of cercarial swimming using purely viscous or Stokes flow theories which correspond to the limit $Re =0$ \cite{lealbook}.

To achieve locomotion, cercariae, like any swimmer dominated by viscosity, must change their shape in a manner that breaks time-reversal symmetry \cite{Purcell1977,lauga_powers}. Using image analysis we extracted the dynamic shape of the organism (Fig. \ref{fig:figure_2} a) and observed that their periodic gait consists of successive power and recovery strokes that break time-reversal symmetry. During a power stroke, the tail bends from a straightened state to its maximally curved state while the fork sweeps out an arc, moving almost perpendicular to its longitudinal axis (sequence $1 \rightarrow 2$ in Fig. \ref{fig:figure_2}a, c, d and f). The fork's motion is directed towards the head of the organism, generating a net thrust force that pulls the head forward along the mean swimming direction $\boldsymbol{\hat{s}}$ (sequence $1 \rightarrow 2$ in Fig. \ref{fig:figure_2} a, c). During a recovery stroke the tail returns to its straightened state and the fork moves parallel to its own longitudinal axis (sequence $2 \rightarrow 3$ in Fig. \ref{fig:figure_2} a, c, e and f) pushing the head backwards. Owing to the different orientation of the fork, this backward displacement is smaller than that during the power stroke thus leading to a cumulative displacement of the head along $\boldsymbol{\hat{s}}$. The head of the organism additionally generates thrust by breaking time-reversal-symmetry (sequence 1 and 3 in Fig. \ref{fig:figure_2}c and f) in a manner similar to conventional Stokesian swimming strategies which involve waving an elastic arm \cite{wiggins1998}. 

At low $Re$ the fluid drag on a body is only weakly anisotropic as compared to the same body at high $Re$. For slender bodies, such as the fork in cercariae (aspect ratio $\approx 20$), the drag at low $Re$ is around twice as large when the body moves perpendicular to its longitudinal axis as compared to parallel to this axis \cite{cox1970sbt,batchelor_1970}. By moving the fork perpendicular and parallel to its axis during the power and recovery strokes, respectively, cercariae seem to maximally utilize this weak anisotropy. 

The kinematics of the fork and head in relation to the tail, as quantified by the respective joint angles (Fig. \ref{fig:figure_2}b), highlight the non-reciprocal nature of the swimming gait which breaks time-reversal symmetry (Fig. \ref{fig:figure_2}f). Interestingly, the tail bends symmetrically and does not contribute to the breaking of time-reversal symmetry (right axis of Fig. \ref{fig:figure_2} c). Therefore cercarial locomotion crucially depends on the degree of freedom at the tail-fork and tail-head joints (Fig. \ref{fig:figure_2}b). This degree of freedom appears to be a result of increased local flexibility in the joints due to smooth-muscle mediated constrictions in the transverse dimension of the organism \cite{Graefe1967,Reger1976}. 

\section*{Theoretical models for swimming cercariae}
\label{sec:model}
\begin{figure}
\begin{center}
\includegraphics[width=0.8\textwidth]{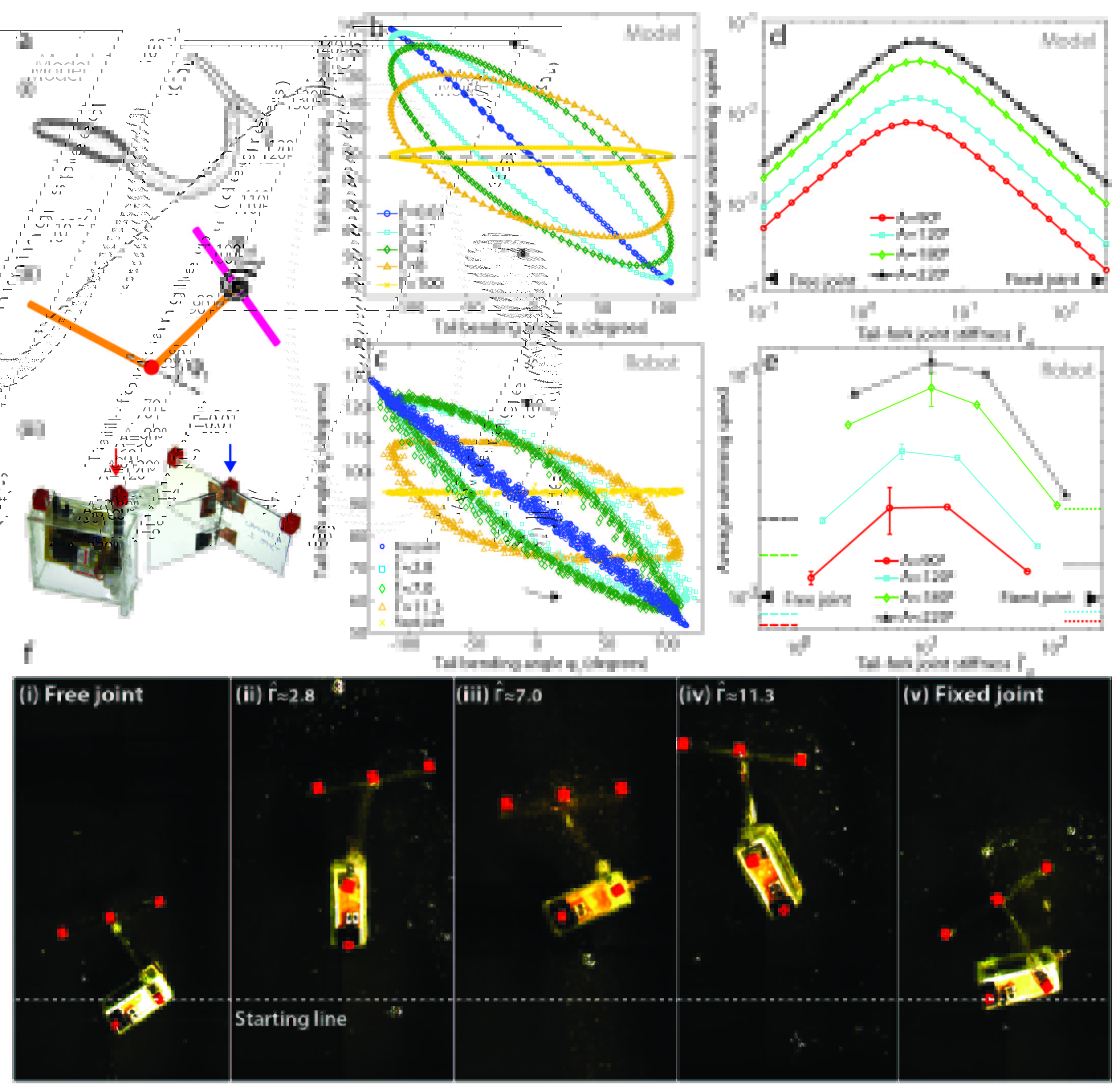} 
\caption{\label{fig:figure_3} \textbf{Theoretical `T-swimmer' and scaled-up robotic model for swimming \textit{S. mansoni} cercariae} \textbf{a,} Schematic of (i) cercariae swimming tail-first and (ii) proposed `T-swimmer' model. (iii) Photograph of a macroscale, self-propelled, `T-swimmer' robot (Methods) designed to swim in a high-viscosity fluid (corn syrup, viscosity $\mu \approx 8 ~Pa ~ s$) chosen to be dynamically similar ($Re_{robot} \approx 0.2$) to swimming cercariae. The joint between the longitudinal links (red dot in (ii)) is active and actuated periodically with a given amplitude $A$ and frequency $f$. The transverse joint is assumed to be a passive linear torsional spring (depicted as a black spiral) with stiffness $\Gamma_{tf}$ to model the flexibility of the tail-fork joint in cercariae. The red and blue arrows in (iii) indicate the active and passive joints, respectively in the robot. \textbf{b, c} Phase plots of the joint angles $\phi_{tf}$ and $\phi_t$ for a T-swimmer model (\textbf{b}) and robot (\textbf{c}) for a range of $\hat{\Gamma}_{tf}$ showing the non-reciprocal nature of the swimming cycle, where $\hat{\Gamma}_{tf}$ is $\Gamma_{tf}$ normalized by the torque scale $\mu f l_c^3$. The arrows indicate the direction of phase trajectories. \textbf{d, e} Plots of the average swimming speed for the T-swimmer model (\textbf{d}) and robot (\textbf{e}) as a function of $\hat{\Gamma}_{tf}$ for different actuation amplitudes $A$. Both the model (\textbf{d}) and robot (\textbf{e}) swim with the passive joint preceding the active joint, and the average swimming speed shows a single maximum at an $O(1)$ value of $\hat{\Gamma}_{tf}$, highlighting an optimal value of torsional stiffness for a given `T-swimmer'. The horizontal lines in \textbf{e} indicate measured swimming speeds for a robot for a free-joint (dashed lines) and fixed-joint (dotted lines) which are an order of magnitude smaller than the peak values. Error bars correspond to standard-deviations over different experiments. \textbf{f,} Snapshots of final positions after $60$ s of swimming for T-swimmer robots with a range of $\hat{\Gamma}$ (i)-(v) and frequency maintained at $\approx 0.4 Hz$. The white dashed line denotes the starting point of the robots. The free (i) and fixed joint (ii) robots show relatively small displacements.}
\end{center}
\end{figure}
To further understand the swimming mechanism and exactly how these joint angles evolve over time, we hypothesize the simplest passive control strategy where a balance between elastic bending stiffness and viscous drag forces dictates the exact kinematics of cercarial swimming --- an elastohydrodynamic effect, rather than active muscular control. We conjecture that cercariae maintain a fixed local stiffness at these joints using smooth musculature, only modifying the joint stiffness when they change their swimming mode. Our hypotheses are supported by earlier studies, which demonstrate that the striated muscles that are capable of rapid beating are confined to the tail, and not present in the fork \cite{nuttman1975structure,nuttman1974fine}. To test our hypotheses we developed a theoretical model: a `T-swimmer', consisting of three, linear, rigid links (Fig. \ref{fig:figure_3}a (ii)). The first two links correspond to the fore and aft ends of the head-tail portion in cercariae, and third link which is attached transversely at its mid-point, corresponds to the fork (Fig. \ref{fig:figure_3}a (i) and (ii)). In our geometry, as a starting point, we neglect the head in cercariae and also the effects of gravity, to make the simplest possible swimming model (Fig. \ref{fig:figure_3}a (ii)). As the reader will quickly notice, our `T-swimmer' is inspired by Purcell's three-link-swimmer \cite{Purcell1977}, often referred to as the simplest low-Reynolds-number swimmer since it requires only two actively actuated joints. In contrast to Purcell's swimmer, in the T-swimmer only a single longitudinal joint is actively actuated (red dot in Fig. \ref{fig:figure_3}a (ii)), to model the bending of the tail of cercariae, while the transverse joint is passive and assumed to be a linear torsional spring (blue spiral in Fig. \ref{fig:figure_3}a (ii)). 

The parameters of the T-swimmer include the lengths of the links ($l_1,~ l_2, ~ l_3$) and their transverse dimensions ($r_1,~ r_2, ~ r_3$); the actuation amplitude of the active joint $A$, the driving frequency $f$ and the torsional stiffness of the passive joint $\Gamma$ (Fig. \ref{fig:figure_3}a (ii)). We non-dimensionalized the system using the half-length of the organism $l_c = L/2$ and the time scale $t_c = 1/f$. The force and torque scales follow the viscous scaling $F_c = \mu f l_c^2$ and $\tau_c = \mu f l_c^3$, respectively \cite{lealbook}. Therefore the key dynamical parameter of the system is the dimensionless torsional stiffness $\hat{\Gamma} = \Gamma / (\mu f l_c^3)$. 

To specify the hydrodynamic forces and torques on the T-swimmer we used local slender-body-theory \cite{cox1970sbt,batchelor_1970} and solved the resulting equations of motion numerically (Supplementary Information section \ref{sec:theory}). Our results show that the T-swimmer swims with the passive joint preceding the active one, similar to cercariae swimming tail-first. In contrast, a slight variation of the classic Purcell's swimmer, which has one active and one passive joint, swims in an opposite direction to the T-swimmer, with the active joint preceding the passive one \cite{passov2012dynamics}. Our results demonstrate that a simple transition from a longitudinal to a transverse link, either of which are attached via a torsional spring, results in a reversal in swimming direction - a viable strategy for any organism trying to reverse swimming directions from head-first to tail-first.

For validating our theory we also performed scaled-up experiments on a centimetre-scale ($L_{robot} \approx 10 ~ cm$) robot mimicking our T-swimmer (Fig. \ref{fig:figure_3}a (iii)). The robot was immersed in a corn-syrup medium to achieve dynamically similar conditions ($Re_{robot} \approx 0.2$) to swimming cercariae (Methods). We found that both the T-swimmer (Fig. \ref{fig:figure_3}b) and the robot (Fig. \ref{fig:figure_3}c), reproduce the non-reciprocal swimming gait of cercariae thereby supporting our hypothesis (Fig. \ref{fig:figure_2}f). For both the T-swimmer and the robot, we varied $\hat{\Gamma}$ and $A$ and found that in the upper and lower limits of joint stiffness ($\hat{\Gamma} \gg 1$ and $\hat{\Gamma} \ll 1$), the gait becomes reciprocal (Fig. \ref{fig:figure_3}b, c) and the swimming speed averaged over $> 60$ cycles is negligible (Fig. \ref{fig:figure_3}d, e, f). This is explained by the fact that for $\hat{\Gamma} \gg 1$ the swimmer has a single degree of freedom since the tail-fork joint is rigid, and for $\hat{\Gamma} \ll 1$, there is again a single effective degree of freedom due to the torque free condition at the tail-fork joint. For both the T-swimmer and the robot, there is an intermediate value of torsional stiffness $\hat{\Gamma}$ that optimizes the average swimming speed (Fig. \ref{fig:figure_3}d, e and f). This happens due to resonant interaction between the driving time scale ($t_{forcing}=1/f$) and the intrinsic relaxation time of the torsional spring ($t_{relaxation}=\mu l_c^3/\Gamma$) i.e $1 / f \sim \mu l_c^3/\Gamma$. 

\section*{Comparison of models and biological experiments}
\label{sec:model_expmt}
\begin{figure}
\begin{center}
\includegraphics[width=0.8\textwidth]{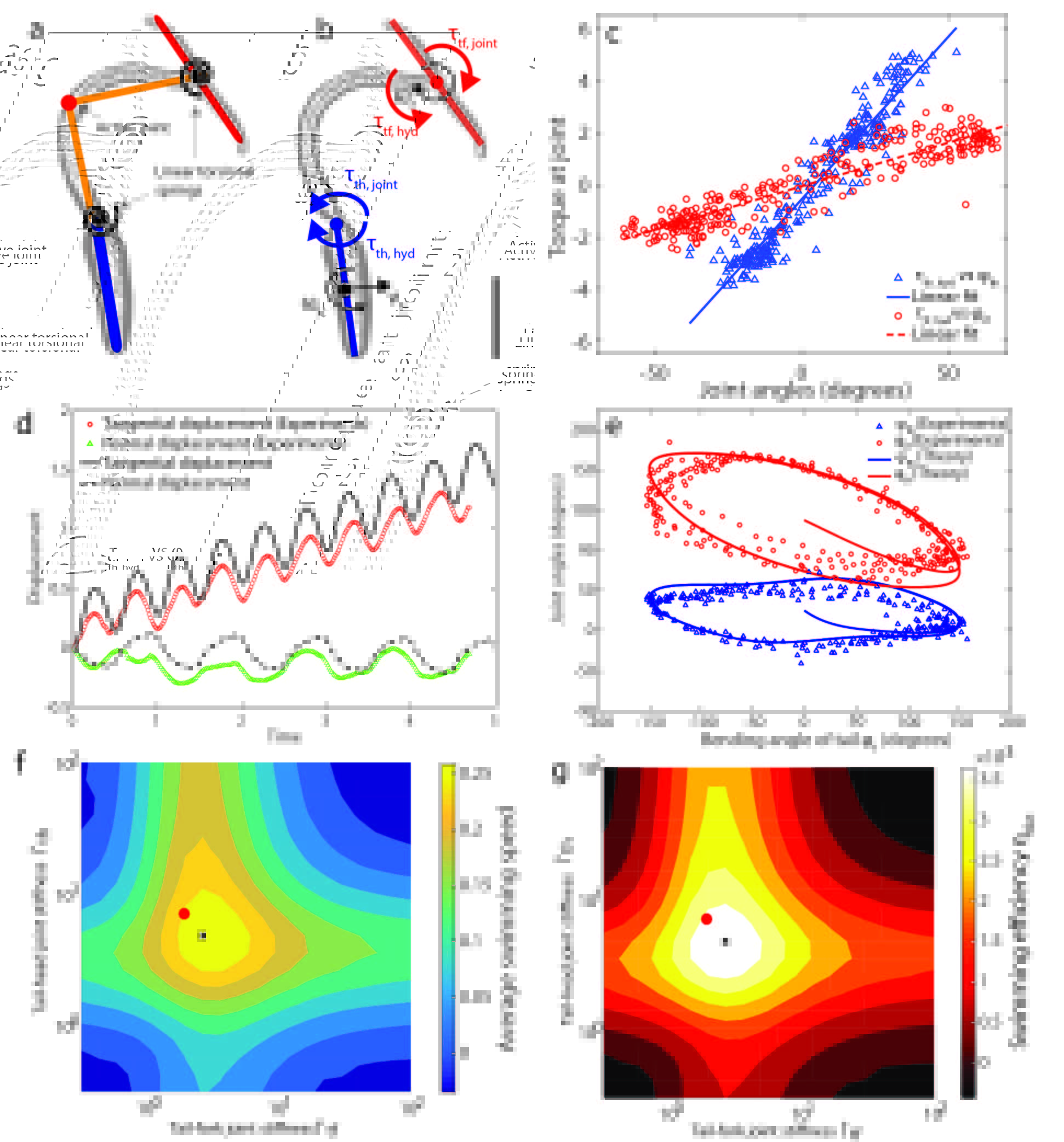}
\caption{\label{fig:figure_4} \textbf{Comparison between four-link `T-swimmer' and swimming \textit{S. mansoni} cercariae.}\textbf{a,} Schematic of a four-link swimmer model overlayed on an image of a cercariae. The flexibility at the tail-fork and tail-head joints in cercariae is modelled via linear torsional springs depicted as black spirals at the respective joints. \textbf{b,} Free-body-diagrams of the head and fork showing translational and angular velocities and the resulting hydrodynamic torque about the respective joints. This torque is balanced instantaneously by the mechanical torques at the joints, thus allowing \textit{in situ} means of estimating these torques (Methods and Supplementary Information section \ref{sec:bending}). Scale bar $100 \mu m$. \textbf{c,} Plot of the estimated torques at the tail-head (blue triangles) and tail-fork (red circles) joints as a function of the corresponding joint angles $\phi_{th}$ and $\phi_{tf}$. The slope of the respective linear fits (R-squared tail-head $ 0.7$ and tail-fork $ 0.8$) shown as solid blue and dashed red lines gives the joint stiffnesses ($\hat{\Gamma}_{th}$ and $\hat{\Gamma}_{tf}$) in live, swimming cercariae. \textbf{d,} Time series plot of displacements of the head along $\boldsymbol{\hat{s}}$ and $\boldsymbol{\hat{n}}$ for the four-link T-swimmer with estimated joint stiffnesses (black solid and dashed curves) compared with experimental measurements (red circles and green triangles). \textbf{e,} Phase plot between $\phi_{th},~ \phi_{tf}$ and $\phi_t$ for the four-link T-swimmer, shown as blue and red lines, and for cercariae shown as blue triangles and red circles. \textbf{f, g} Contour plots of average swimming speed (\textbf{f}) and swimming efficiency (\textbf{g}) as a function of the joint stiffnesses $\hat{\Gamma}_{tf}$ and $\hat{\Gamma}_{th}$ show a single optimal point (black crosses). The red dots indicate the values estimated for cercariae showing that they optimize their joint stiffnesses to swim efficiently.
}
\end{center}
\end{figure}

As a next extension to the three-link T-swimmer, we added a fourth link to account for the thrust generated by the head in cercariae and also included the effects of gravity (Supplementary Information section \ref{sec:cog}). This fourth link is attached longitudinally via a torsional spring to model the flexibility of the tail-head joint (Fig. \ref{fig:figure_4}a). The resulting model can now be compared to experimental measurements on \textit{S. mansoni} cercariae.

To make this comparsion, we need the torsional spring stiffness of the two joints ($\hat{\Gamma}_{th}$ and $\hat{\Gamma}_{tf}$) in a living (swimming) cercariae. We developed a novel \textit{in situ} measurement using high-speed imaging data (Fig. \ref{fig:figure_2}) and hydrodynamic theory. To do this, first we used slender-body-theory \cite{lealbook} to estimate the instantaneous torques at the two joints $\boldsymbol{\tau}_{tf}$ and $\boldsymbol{\tau}_{th}$ from the translational and angular velocities of the fork and head (Fig. \ref{fig:figure_4}b and Supplementary Information section \ref{sec:bending}). We then investigated the relationship between these torques and the corresponding joint angles $\phi_{tf}$ and $\phi_{th}$ and found that they are linearly correlated (Fig. \ref{fig:figure_4}c). This linearity implies that the two joints are indeed well-approximated as linear torsional springs, serving as \textit{a posteriori} verification of our T-swimmer model. Finally we obtained the torsional spring stiffness of the two joints as the slopes of the linear fits. 

By comparing our four-link T-swimmer with experiments on \textit{S. mansoni} cercariae, we find that the theoretical four-link T swimmer indeed captures the swimming displacement per cycle (Fig. \ref{fig:figure_4}d) and the time evolution of the non-reciprocal joint angle kinematics very well (Fig. \ref{fig:figure_4}e). A crucial connection to disease transmission lies in the swimming efficiency of this human parasite. Next, we estimate the cercarial swimming efficiency using our T-swimmer model. To measure how effectively cercariae can swim up the water column, we defined this efficiency as the ratio of average distance covered relative to the average viscous power dissipation in the fluid during a swimming cycle $\eta_{dist} = X_{stroke}/ P_{diss, stroke}$ (Supplementary Information section \ref{sec:theory}). By exploring a range of values in the $(\hat{\Gamma}_{tf},\hat{\Gamma}_{th})$ plane we find a single optimal point for both the swimming speed (Fig. \ref{fig:figure_4} f) and efficiency (Fig. \ref{fig:figure_4} g). Interestingly, we observed that the joint spring stiffnesses estimated for cercariae (red dots in Fig. \ref{fig:figure_4} f and g) lies very close to this optimal peak. This suggests that cercariae maintain the stiffness of these two joints at the optimal range required for efficient swimming. This tuning may have resulted from strong evolutionary pressures on the parasite since the host-seeking processes is energy constrained \cite{lawson1980survival,Whitfield2003}.  

\section*{Outlook}
Our work points to an unusual elastohydrodynamic effect in cercarial swimming, where the tail provides the energy for propulsion but no thrust while tail-head and tail-fork joints act as passive torsional springs providing all the thrust. This passive control strategy is minimalistic in nature where the only control lies in tuning the flexibility of the respective joints.  Furthermore, cercariae are the first reported organisms to use both a longitudinal (the head) and transverse element (the fork) \cite{ganesh_2012_review,brennen1975locomotion} to generate a net propulsion from bending waves propagating in opposite directions (due to the tail).

Since current mass drug administration strategies have significant limitations \cite{cioli2014schistosomiasis}, mechanistically linking biophysics of human-seeking parasites such as \textit{S. mansoni} to disease transmission in ecological field conditions could inspire a new paradigm of environmental control of this neglected tropical disease. The limited energy reserves of the cercarial stage of this parasite might be its Achilles heel. Being of a higher density than water, simply targeting the swimming motility could impair the parasites ability to aggregate near the air-water interface, thereby reducing transmission rates. Towards this, our study of the swimming mechanics has given the first quantitative estimate of the swimming efficiency, and allows an understanding of how perturbations to various parameters affect the motility of cercariae. 
 \section*{Acknowledgements}
 We thank all members of Prakash lab for fruitful discussions. DK is supported by a Stanford Bio-X Bowes fellowship. GK was supported by scholarships from the Onassis and the A.G. Leventis Foundation. MP is supported by a Multidisciplinary University Research Initiatives (MURI) grant. We thank Dr. Judy Sakanari of UCSF for providing lab space and organisms.  We thank Mattias Lanas for the scientific illustrations of cercariae in their natural habitat. 


\newpage
\newpage
\makeatletter
\renewcommand{\thefigure}{S\arabic{figure}}
\makeatother
\setcounter{figure}{0}
\setcounter{section}{0}
\resetlinenumber[1]
\newpage
\begin{center}
\Large{\textbf{Supplementary Information} for ``\textit{Schistosoma mansoni} cercariae exploit an elastohydrodynamic coupling to swim efficiently''}
\end{center}
\section{Methods}
\label{sec:methods}

\subsection{Experiments on Live Cercariae}
\subsubsection{Organism preparation} 
\label{ssub:organism_preparation}

Live organisms were obtained from UCSF (Judy Sakanari Lab) and experiments were conducted there. Experiments on vertical swimming of cercariae were carried out in a flow chamber with height 4 cm, width 1 cm and thickness 1.4 mm. Standard soft lithography techniques were used to create Polydimethysiloxane (PDMS) chambers of the required dimensions which were then bonded on to corona treated glass slides (Electron Microscopy Sciences) of dimensions ($75 mm \times 51 mm \times 1.2 mm$). One side of the resulting flow chamber was left exposed to air so that when the chamber was partially filled, an air-water interface was present. This setup mimics the natural conditions in which cercariae swim vertically in the water column. Cercariae freshly shed from a snail were washed and a fixed volume of snail-conditioned-water ($100 \mu l$) containing the organism was introduced into the chamber so that around $75 \%$ of the chamber height was full. 

\subsubsection{Imaging system} 
\label{ssub:imaging_system}
For imaging the swimming of cercariae, the flow chamber was mounted statically on a optics board( Thor Labs) with a plane mirror as a backing. A reflection microscopy platform was constructed using a modular infinity microscope system (Applied Scientific Intrumentation), along with warm white light (500 mW LED source ,Thor Labs, MWHCL3) channeled through a fibre optic cable and combined with a collimator lens and beam-splitter (Edmund Optics, 50R/50T plate). Infinity-corrected, long-working-distance objectives (Mitutoyo Corporation) were used to achieve a magnification range of $1 \times$ to $50 \times$, with a corresponding resolving power of $11 \mu m$ to $0.5 \mu m$. Image acquisition, at up to 2000 fps, was done through a high-speed camera (Phantom v12.10). The entire reflection microscopy system including the high-speed camera was mounted on a $x-y-z$ stage to provide translation capability along all three axes.

\subsubsection{Swimming trajectory and Particle-Image-Velocimetry} 
\label{ssub:swimming_trajectory_and_particle_image_velocimetry}
For obtaining bulk-swimming trajectories, measurements were made using the reflection microscopy system at $100$ fps at $2\times$ magnification. This allowed a large enough field-of-view (circular with diameter $12 mm$, $\approx 30$ body lengths) so that bulk swimming characteristics could be observed yet the resolution allowed detection of the precise shape of individual cercariae. The resolution of the shape of the organism was important to classify the trajectories into different swimming modes.

For detailed analysis of the swimming kinematics of individual organisms, data was obtained at $1000$ fps at $10 \times$ and $20 \times$ magnification. This allowed a swimming stroke to be resolved into $\approx 50$ time points which was enough to resolve the detailed motions of the head, tail and fork during a swimming stroke, which were obtained using an in-house image processing tool. For measuring the flow-field around the organisms, the snail-conditioned-water was seeded with $2 ~ \mu m$ beads (Fluoresbrite, calibration grade) and images were acquired at $1000$ fps at $10 \times$ and $20 \times$ magnification. From these images the shape of the organism was extracted using an in-house image processing tool and applied as a dynamic mask to exclude regions containing the organism. Velocity fields were obtained from this post-processed image using an open-source Particle-Image-Velocimetry tool implemented in MATLAB \cite{thielicke2010pivlab}. 

\subsection{Scaled-up experiments}
An autonomous macro-scale robot was designed and built to fully explore the novel `T-swimmer' geometry proposed in the paper. The robot had three links in the form of rectangular plates fabricated from acrylic sheets of thickness $1.5 ~ mm$. Of these two were longitudinal links while one was transverse resulting in a three-link `T-swimmer'. The height of the links was $5 ~cm$ and the total length of the robot was $\approx 10 ~ cm$. The joint between the longitudinal links is actively driven by a micro-metal gear motor (100:1 medium-power with dual-shaft, Pololu Robotics and Electronics) controlled by circuits and power systems which are onboard the robot to avoid the undesirable effects of external wires or control lines. A closed-loop control circuit was developed using a \textit{LB 1836 IC motor driver, Texas Instruments} functioning as a half H-bridge to change the direction of rotation of the motor. To precisely measure the angular displacement of the motor shaft an optical encoder system (Pololu Robotics and Electronics) was attached to the auxiliary shaft of the motor and its output fed to a differential comparator (\textit{LM 311, Texas Instruments}). This resulted in an angular resolution of $0.72 \degree$. The motors and control circuit were interfaced using a \textit{PICAXE 08M2} micro-controller which allowed programmability of the amplitude of the active joint. The motor and control circuits were independently powered by 3.7 V Lithium-Polymer batteries \textit{(Tenergy Corporation)}. The motor, control electronics and batteries were housed in a water-tight box in the distal longitudinal link of the swimmer.

The stiffness of the passive `T-joint' was modified by using torsional springs ($180 \degree$ steel music wire, McMaster-Carr). By using a number of springs in parallel and also by using individual springs of varying stiffness, around two orders of magnitude in torsional stiffness of the passive joint was achieved. 

The `T-swimmer' robot was placed in a tank of dimensions ($90 cm \times 45 cm \times 40 cm$) filled with light corn syrup (Karo). The rheological properties of the corn syrup were measured using a high-resolution rheometer (\textit{TA Instrument ARES-G2}) and was found to be Newtonian over a shear rate of $0.1 ~ - ~ 10 ~ s^{-1}$ with a viscosity of $\mu \approx 8 Pa s$ at $20 \degree C$. The robot was made neutrally buoyant and stable by suitably adding ballast so that the centre-of-gravity was close to the centre-of-buoyancy. The robot thus swims in a horizontal plane and gravity does not play a role in the motion. The microcontroller system was pre-programmed and the robot was placed near the middle of the tank. An overhead camera (Canon EOS Rebel T5) was set up to record the motion of the robot. To capture the shape kinematics colored markers were placed at different points on the robot (Fig. \ref{fig:figure_3} f). The resulting videos (Fig. \ref{fig:figure_3} f) were processed to obtain the $x-y$ locations of the markers. This allowed us to derive the both the overall kinematics of the robot as well as the kinematics of the joint angles (as shown in Fig. \ref{fig:figure_3} c, e).

\subsection{Image Processing} 
\label{ssub:image_processing}
For the analysis of the grayscale video data from the experiments with live cercarie, two codes were used.  The first code was used to track the trajectories of multiple cercarie swimming in the vertical chambers (Fig. \ref{fig:figure_1} c). This code was written in MATLAB and utilized an open-source code (MATLAB Particle Tracking Code by Daniel Blair and Eric Dufresne) to track the cercarie. The post-processing of the tracking data, including the identification of swimming modes, was done using a custom MATLAB code. To identify the swimming mode for each tracked cercarie, the instantaneous velocity was calculated and the swimming mode was determined based on the sign and the magnitude of that velocity. The second code was used to study single cercarie (Fig. \ref{fig:figure_2}) by identifying their distinct parts, including the head, the fork, and the tail, in order to track their positions and shapes as cercarie swim over time. The head and the fork were tracked by a user-defined mask that was manually set for the first frame of each video sequence. For the subsequent frames the head and the fork were tracked automatically using a brightness threshold. The code was written in MATLAB and fit ellipsoids on the head and the fork to extract the coordinates of their centroids, their major and minor axes and the orientation angles for each frame. The tail of the cercarie was extracted by subtracting the head and the fork from the whole body of the organism that was also tracked automatically using a brightness threshold. To extract the curvature and the intersection of the angles of the tail at joints with the head and fork, the tail was skeletonized and a spline was fit along its length. For the analysis of the RGB video data from the scaled-up experiments with the macroscopic robot, a custom code was written in MATLAB. The code used the K-means method to obtain the locations and extract the coordinates of the coloured markers placed at different points of the robot.
\section{Additional Discussion}
\label{sec:fluid_regime}
\subsection{Fluid Dynamical Regime for Cercariae}
\begin{figure}[h!]
\begin{center}
\subfloat[]{\includegraphics[width=0.5\textwidth]{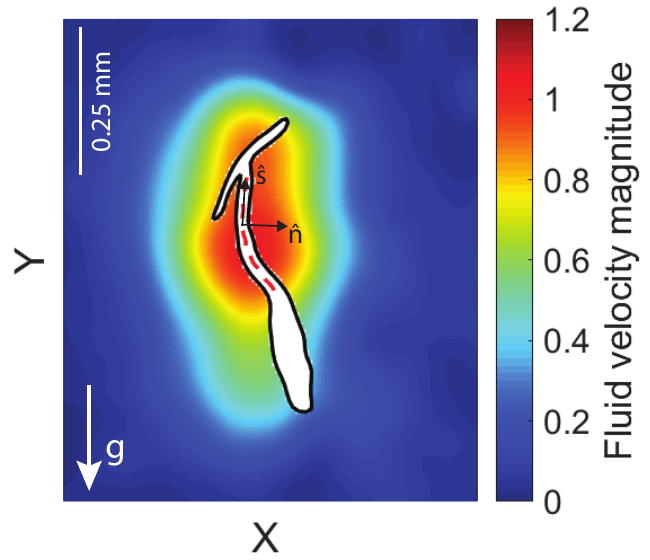} \label{fig:SI_figure_1a} } \\
\subfloat[]{\includegraphics[width=0.465\textwidth]{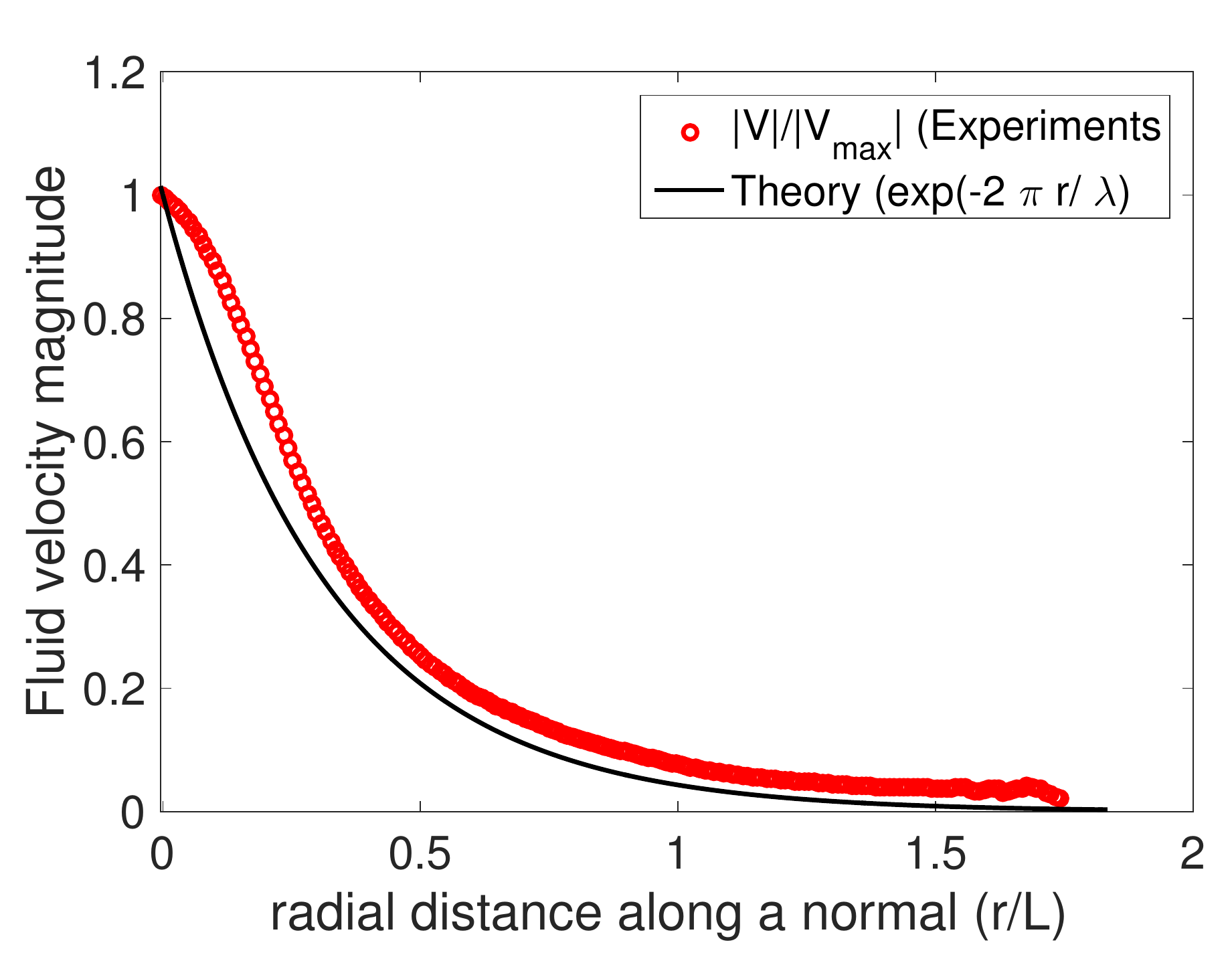} \label{fig:SI_figure_1b} }
\subfloat[]{\includegraphics[width=0.495\textwidth]{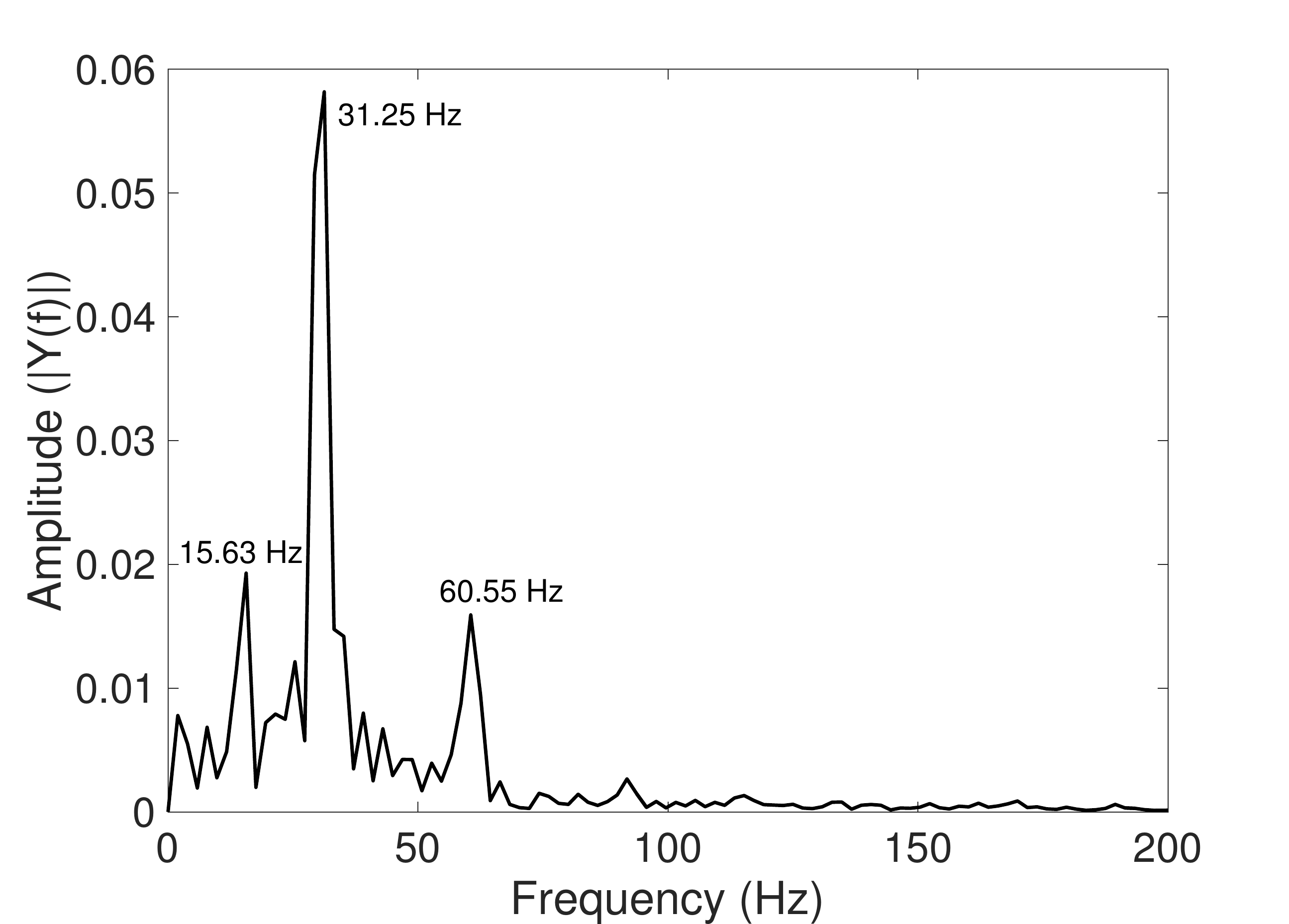} \label{fig:SI_figure_1c} }
\caption{\label{fig:SI_figure_1} \textbf{a,} Fluid velocity magnitude (normalized by the velocity scale $l_c f$) in a region surrounding a freely swimming cercaria. \textbf{b,} Decay of the fluid velocity magnitude normal to the axis of the tail ($\boldsymbol{\hat{n}}$), ensemble averaged over different spatial locations along the tail and time instants, plotted as a function of normalized distance away from the swimmer surface ($r/L$), where $L$ is the length of the organism. $r=0$ corresponds to the surface of the swimmer. The solid black curve is the theoretical prediction for an undulatory swimmer at low-Reynolds-numbers \cite{lighthill1976}. \textbf{c,} Fourier transform in time of the average fluid velocity magnitude in a region surrounding the swimmer. The extent of this region is around $2$ swimmer lengths.
}
\end{center}
\end{figure}
In this section we establish that the flow around swimming cercariae, and hence the swimming mechanics, is dominated by viscous rather than inertial effects. To discuss the hydrodynamics of the ambient fluid, the relevant dimensionless number is the Reynolds number ($Re$), which is defined as:
\begin{equation}
 Re = \frac{U L}{\nu} ,
\end{equation}
where $U$ is the swimming speed of the organism, $L$ is the characteristic length scale, and $\nu = 10^{-6} m^2 s^{-1}$ the kinematic viscosity of water which is the ambient fluid. Low-Reynolds-numbers (or Stokesian) flows are characterized by the relative dominance of viscous stresses compared to inertial stresses in the flow. For cercariae, the Reynolds number based on the \textit{maximum} measured swimming speed of $1.5 mm/s$ and \textit{maximum} half length of the organism $L= 250 \mu m$ (from the tip of the head to the base of the fork, with the main tail fully extended), is around $0.3$. In the near-field of the organism, a second Reynolds number may be defined based on the beating frequency of the tail as:
\begin{equation}
Re_{f} = \frac{f L d}{\nu}
\end{equation}
where $f$ is the beat frequency of the tail whose maximum value from experiments is around $20 Hz$, and $d$ is the tail diameter which is measured to be $20 \mu m$. This Reynolds-number then has the value:
\begin{equation}
Re_{f} = \frac{20 (250 \times 10^{-6}) 20 \times 10^{-6}}{1 \times 10^{-6}} = 0.1
\end{equation}
Both these Reynolds numbers are higher than typically studied Stokesian micro-swimmers such as bacteria and algae which is around $Re \sim \mathcal{O}(10^{-4})$\cite{lauga_powers}, and fall near the upper-bound of what is usually regarded the Stokesian regime. Hence it is worth investigating in detail if the flow-field and hence swimming mechanics is qualitatively dominated by viscous or inertial effects. To do so we consider several metrics based on the measured velocity field around freely swimming cercariae.

 We first consider the spatial decay of the fluid velocity magnitude in a direction normal to the swimmer surface (see Fig. \ref{fig:SI_figure_1a}), ensemble averaged over different time and spatial locations along the tail of the swimmer (Fig. \ref{fig:SI_figure_1b}). The values are normalized by the mean of the maximum fluid velocity magnitude for each time instant and spatial location. The resulting data is seen to collapse onto a single curve which compares well to an exponential decay of the form $exp(-2 \pi r/\lambda)$, where $r$ is the spatial distance away from the swimmer surface and $\lambda$ is the characteristic wavelength of the undulations, which for the case of cercariae is given by $\lambda = 2 l_{tail}$. This is because the tail bends in a symmetric fashion with the tail-fork and tail-head joints acting like nodes. This exponential decay is a characteristic feature of Stokesian flow around an undulatory swimmer \cite{sznitman2010propulsive,lighthill1976}.

Fig. \ref{fig:SI_figure_1c} shows the Fourier transform of the spatially-averaged fluid velocity magnitude as a function of time. The fundamental frequency of the signal is seen as a sharp peak at $f \approx 15 Hz$ which is the beat frequency of the organism in this particular experiment, while the higher frequency peaks occur at approximately $2 f $ and $4 f$. This characteristic nature of the power spectrum with sharp peaks at $f$, $2 f$ and $4 f$, points to the flow around the organism adjusting instantaneously to the motion of the organism. This linear response of the fluid to the imposed boundary motion is a typical feature of low-Reynolds-number flows.

The peak at $2 f$ is attributed to the swimming gait consisting of beating on the dorsal and ventral side of the organism. On the other hand, the $4 f$ peak is likely due to the motion of the fork, which, during a swimming stroke, undergoes four distinct motions (two power and two recovery strokes in alternating fashion). That this is due to the presence of the fork is seen by comparing the spectrum for cercariae to that of \textit{C. elegans} \cite{sznitman2010propulsive}. This organism being a simple undulatory swimmer with no fork shows peaks at $f$ and $2 f$ but no peak at $4 f$.  

In summary, all the metrics for the fluid flow surrounding the organism presented in the main paper and also above point to the underlying Stokesian flow physics which governs the swimming mechanics of cercariae. 
\subsection{Estimating the location of the centre-of-gravity and centre-of-buoyancy for cercariae}
\label{sec:cog}
\begin{figure}
\begin{center}
\includegraphics[width=1.0\textwidth]{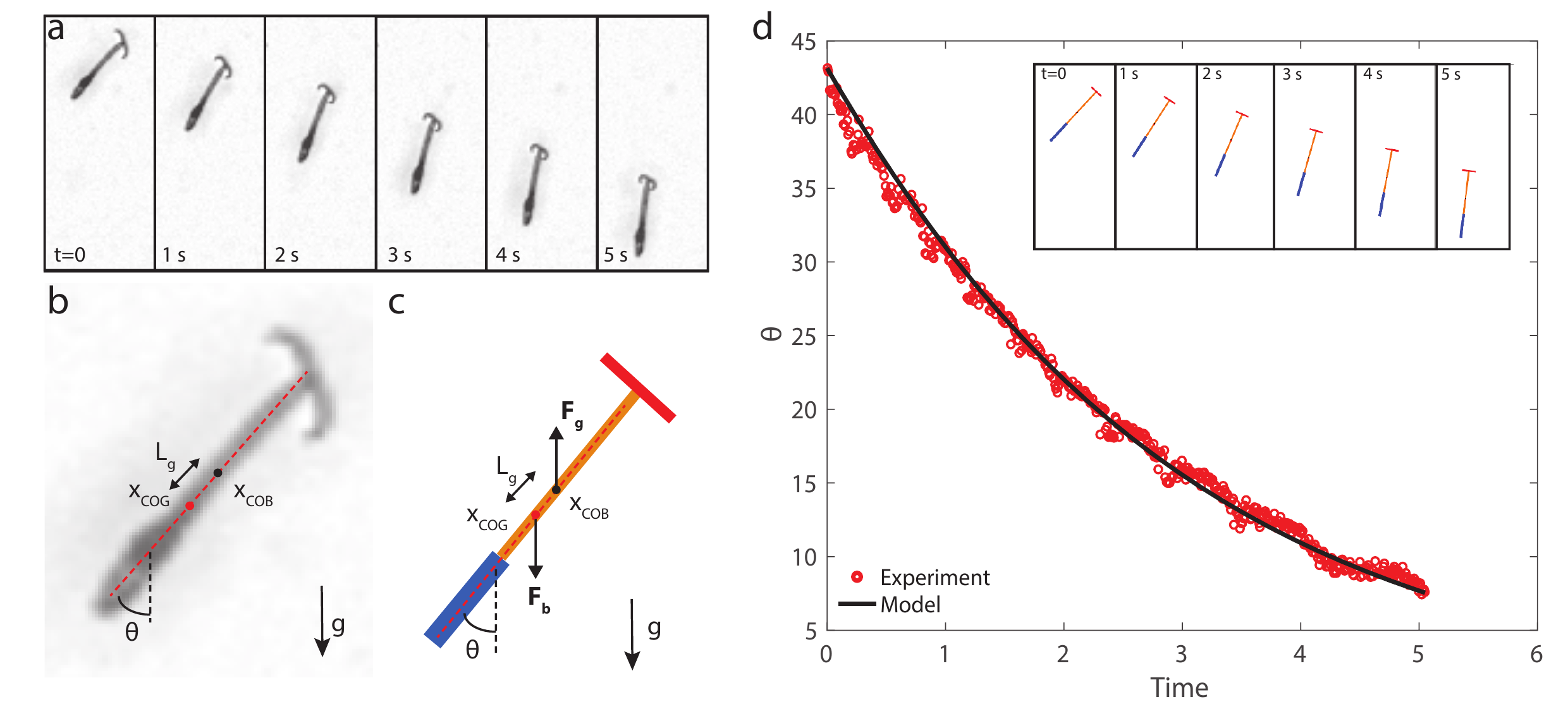} 
\caption{\label{fig:SI_figure_reorient} \textbf{a,} Snapshots at $1$ s intervals of a cercaria passively sinking and reorienting due to gravity. \textbf{b,} A schematic depiction of the location of the center-of-gravity and centre-of-buoyancy both of which, by symmetry, lie on the central axis, separated by a distance $L_g$. \textbf{c,} A reduced-order model to allow calculation of $L_g$ based on a balance of gravitational and viscous torques. \textbf{d,} Experimental measurements and theoretical prediction of angle of the longitudinal axis with the vertical during passive sinking. Inset shows snapshots of the model at $1$ s intervals.  
}
\end{center}
\end{figure}
Cercariae are negatively-buoyant, bottom-heavy swimmers. To model their swimming it is important to know their excess density compared to water and also the realigning torque due to gravity. In this section, we use trajectories of cercariae passively sinking and reorienting due to gravity to estimate this excess density as well as locations of centre-of-gravity (COG) and centre-of-buoyancy (COB) in cercariae. This in turn we use to estimate the gravitational torque.  

Since there is no available information regarding the mass or density of individual anatomical parts of cercariae, we first restrict ourselves to solving for the average density of the organism (defined as mass of organism by total volume) based on the sinking speeds of the cercariae. Further, we assume each part of the cercarial anatomy has a uniform density. This is a reasonable assumption since each of these sub-divisions have a distinct set of muscle groups and smaller scale anatomical structures unique to that portion \cite{Reger1976}. It is possible to show that assumption of uniform density for each segment of the cercaria precludes the effect of gravity in the bending of the tail-fork and tail-head joints. 

The shape of a cercariae sinking passively with the fork partially extended is shown in Fig. \ref{fig:SI_figure_reorient}a, and the reduced-order-model of the same using slender rods is shown in Fig. \ref{fig:SI_figure_reorient}b. In trajectories where the cercaria sank while oriented parallel to gravity, there is no reorientation and we can simply equate the net buoyant force to that due to the total hydrodynamic drag giving:
\begin{equation}
(\rho_{cerc}- \rho_f) V_{cerc} g = \mu U_{sink} (C_{L,head} l_{head} + C_{L,tail} l_{tail} + C_{N,fork} l_{fork})
\end{equation}
where $\rho_{cerc}$ and $\rho_f$ are the densities of the organism and ambient fluid, respectively; $V_{cerc}$ is the total volume of a cercariae while sinking passively estimated from the measured dimensions assuming a circular cross-section throughout its body, $g = 9.81 ~ m ~ s^{-2}$ is the acceleration due to gravity, $U_{sink} = 0.13 ~ mm ~ s^{-1}$ is the average sinking speed (over $>1000$ trajectory segments). The $C_L$ and $C_N$ are the drag coefficients of the different parts of the cercaria which are modeled as slender rods (see Section \ref{sec:theory}). This gives an estimate for the excess density as $\rho_{cerc}- \rho_f \approx 70 ~ kg ~ m^{-3}$.

Next we estimate the location of the COG of the organism. The COB of a submerged object in a fluid of homogeneous density is just the centroid. The location of the COB is thus known given the dimensions of the organism. Let $L_g$ be the distance between this COB and the unknown location of the COG. The reorientation of cercaria happens due to a balance between the gravitational torque $\rho_f V_{cerc} g L_g sin{(\theta)}$, where $\theta$ is the angle made by the longitudinal axis of the organism with the vertical (see Fig. \ref{fig:SI_figure_reorient}b) and the viscous torque. The net viscous torque is computed for the shape in Fig. \ref{fig:SI_figure_reorient}c using the slender-body drag coefficients for the head, tail and fork segments (also see Section \ref{sec:theory}), with the simplification that the swimmer is rigid and passive. The resulting dynamics of the orientation in terms of $\theta$ as a function of time were compared with experimental measurements for different values of $L_g$ and the value that minimized the error between the two was found to be $L_g = 8.3 \pm 1.1 \mu m$ (Fig . \ref{fig:SI_figure_reorient}d).
\subsection{Estimating bending stiffness from high-speed kinematics data}
\label{sec:bending}
\begin{figure}
\begin{center}
\includegraphics[width=0.5\textwidth]{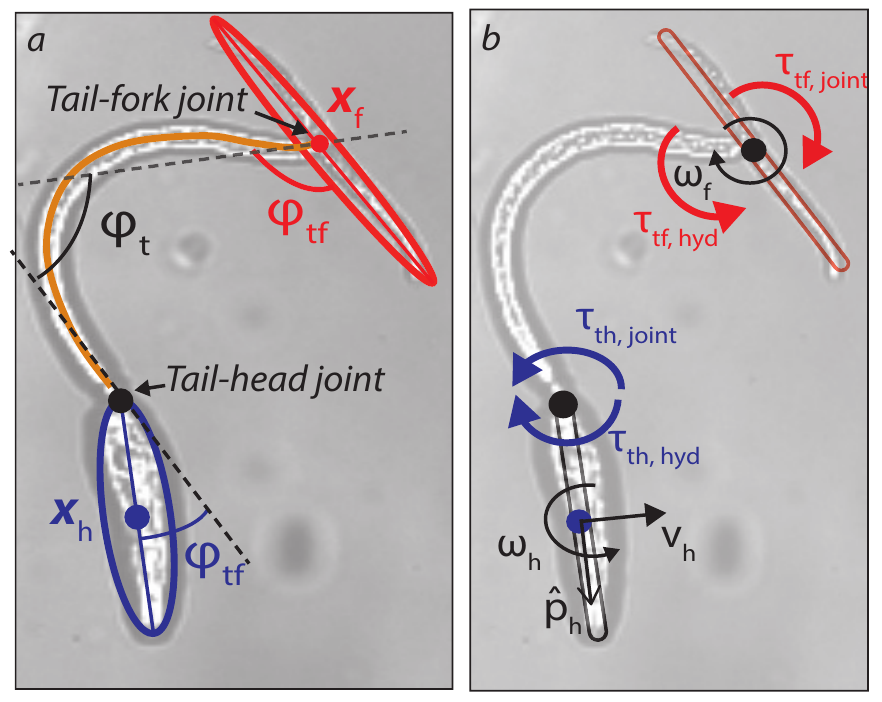} 
\caption{\label{fig:SI_joint_stiffness} \textbf{a,} Snapshot of a cercaria swimming freely with reduced-order representations of the head, tail and fork overlaid. These representations allows position and shape information to be extracted from the high-speed videos. The head (blue) and fork (red) have positions (denoted by blue and red dots) and orientations given by the centroids and major axes of the respective ellipses. The shape of the tail is extracted by fitting a spline curve along its central axis (solid orange line). The bending of the tail $\phi_t$ and the tail-head and tail-fork angles $\phi_{th}$ and $\phi_{tf}$ are also shown. \textbf{b,} Representation of the kinematics of the head and fork along with the hydrodynamic and mechanical torques at the joints. The head and fork are considered to be slender-rods with equivalent aspect ratio. $\boldsymbol{\hat{p}_h}$ is the orientation vector along the major axis of the head.
}
\end{center}
\end{figure}
In this section we provide details of the procedure to estimate the bending stiffness of the tail-head and tail-fork joints in cercariae . High-speed videos of individual cercariae swimming were analysed using an in-house image processing code (Methods). This analysis separated the cercariae into three parts namely the head, tail and fork. The head and fork were extracted by fitting an equivalent ellipse to the respective part of the organism (Fig. \ref{fig:SI_joint_stiffness}a). A spline fit was used to extract the shape of the tail. The two ellipses and the spline gave us all the necessary morphological information about the organism at each time instant. From this data we can extract the relevant angles at the tail-fork and tail-head joints as shown in Fig. \ref{fig:SI_joint_stiffness}a. 

The velocities of the centroid of the head and fork (assumed to be the centroids of the equivalent ellipses) are extracted using a central-difference of their position data. Additionally we calculated the normal component of the velocity $\boldsymbol{v_h}.\boldsymbol{\hat{n}}$ of the head normal to its major axis as shown in Fig. \ref{fig:SI_joint_stiffness}b. We assume that the head and fork are slender bodies of aspect ratio $\epsilon_{h/f} = \frac{l_{h/f}}{r_{h/f}}$, where $l_{h/f}$ and $r_{h/f}$ are the length and semi-minor axis of the head and fork, respectively, giving $\epsilon_h \approx 9$ and $\epsilon_f \approx 19$. The large values of the aspect ratio justifies the use of slender body theory.

 With this assumption we can estimate the net hydrodynamic torque at the tail-head joint as $\boldsymbol{\tau}_{th, hyd} = -C_N |_{h} (l_h^2/2) \boldsymbol{\hat{p}}_{h} \wedge (\boldsymbol{v_h}. \boldsymbol{\hat{n}}) \boldsymbol{\hat{n}} - C_r |_{h} l_h^3 \boldsymbol{\omega_h}$, where $\boldsymbol{\hat{p}}_{h}$ is the orientation unit vector along the major axis of the head (Fig. \ref{fig:SI_joint_stiffness} b). Here $C_N$ is the resistance coefficient for motion of a slender body of length $l$ normal to its length and is given by $C_N = 4 \pi \mu / ln(\epsilon)$, and $C_r$ is the resistance coefficient for solid body rotation about an axis normal to the slender body axis and is given by $C_r = \pi \mu / (3 ln(\epsilon))$, where $\mu$ is the fluid viscosity. Similarly the hydrodynamic torque at the tail-fork joint is give by $\boldsymbol{\tau}_{f, hyd} = - C_r |_{f} l_f^3 \boldsymbol{\omega_f}$. 

 The quasi-steadiness of Stokes flow implies that the instantaneous net torque on the head and fork must vanish \cite{lealbook}. Thus the junction of the head and fork with the tail must supply an equivalent and opposite mechanical bending moment to counterbalance the hydrodynamic torques on the head and fork calculated above. This implies $\boldsymbol{\tau}_{th/tf,hyd} = -\boldsymbol{\tau}_{th/tf, joint}$. Thus we can estimate this bending moment at the joint. Independently, we have measured the bending angles $\phi_{th}$ and $\phi_{tf}$ at the two joints. A phase plot of the calculated torques as a function of the bending angles is given in the main paper in Fig. \ref{fig:figure_4}c. The linear response of the torque to the bending angle points to the elastic nature of the joint and justifies modelling them as linear torsional springs. 
\subsection{Swimmer Kinematics for Horizontally Swimming Cercariae}
\label{sec:hor_swim}
\begin{figure}
\begin{center}
\includegraphics[width=0.8\textwidth]{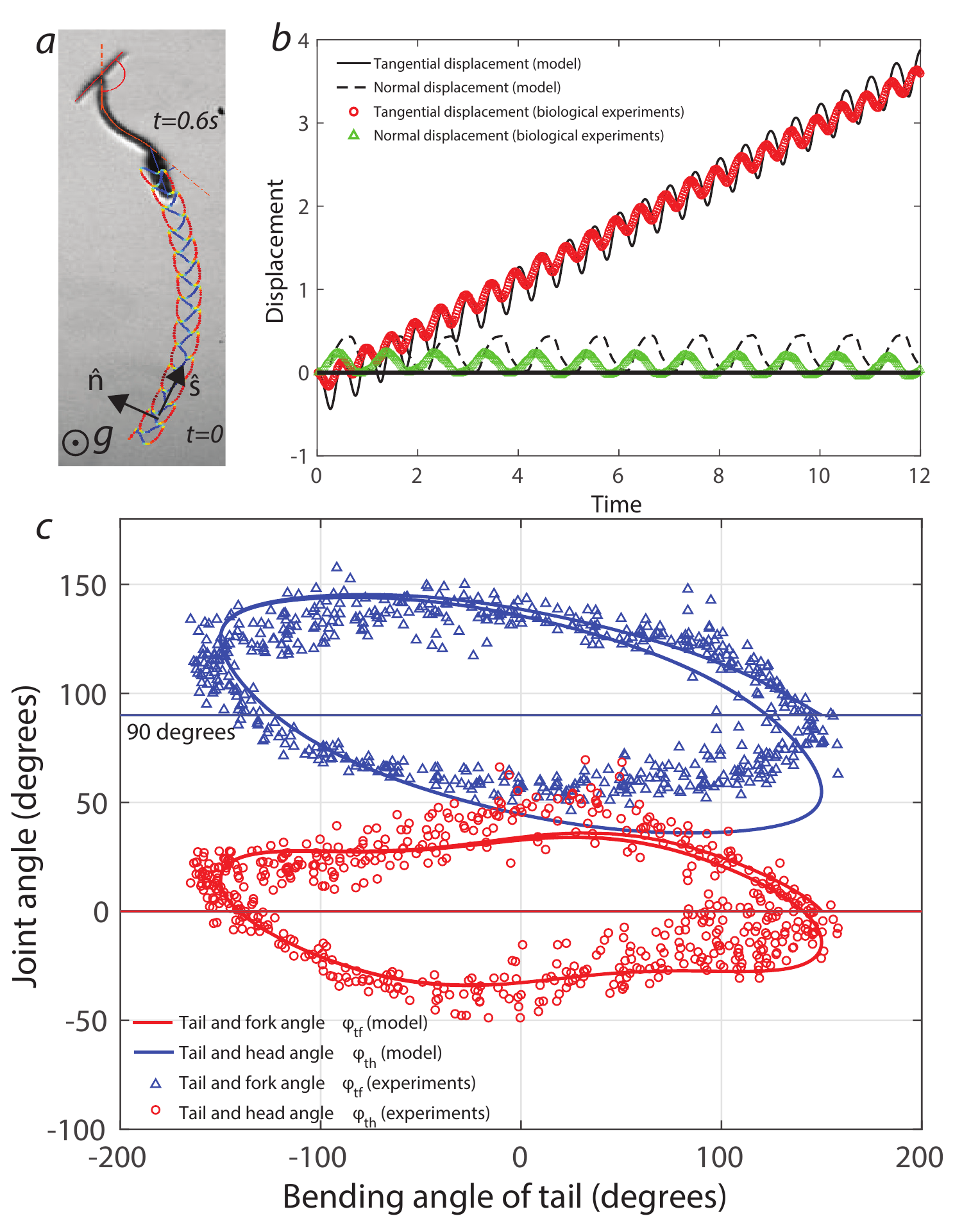} 
\caption{\label{fig:SI_hor_swimming} \textbf{a}, The trajectory of the head of the cercariae swimming in a horizontal plane. The $\boldsymbol{\hat{s}}$ and $\boldsymbol{\hat{n}}$ directions are defined as along the tangent and normal to the local average swimming direction, respectively. \textbf{b}, Displacement of swimmer along $\boldsymbol{\hat{s}}$ and $\boldsymbol{\hat{n}}$. The symbols (red circles and green triangles) represent tangential and normal displacements, respectively, from experimental measurements. The black solid and dashed lines represent tangential and normal displacements, respectively, from the theory. \textbf{c}, Phase plot of the joint angles $\phi_{th}$ and $\phi_{tf}$ as a function of the tail bending angle $\phi_{t}$ for experiments (symbols) and theory (solid lines).
}
\end{center}
\end{figure}
Here we present data on cercariae swimming in the horizontal plane near the interface. This occurs when cercariae continue swimming even after reaching the interface. In this case, gravity is normal to the swimming direction and pointing out of the plane as shown in Fig. \ref{fig:SI_hor_swimming}a (since the imaging is upwards). The kinematics of the organism in terms of the head trajectory and angles at the head and fork joints were found to be similar to the case of vertical swimming (Fig. \ref{fig:figure_2}). Hiwever, the average swimming speed was found to be $3$ body lengths/s which is larger than for the vertical swimming case since gravity is acting in a direction perpendicular to the mean swimming direction. 

Following the method described in Section \ref{sec:bending}, we can estimate the torsional stiffnesses at the two joints in dimensionless terms to be $\hat{\Gamma}_{tf}= 4.2$ and $\hat{\Gamma}_{th}= 8.3$ which are similar to that obtained for the vertical swimming values. Also the average bending angle amplitude of the tail is found to be $A \approx 150 \degree$. Inputting these parameters into the four-link-swimmer model (as described in Section \ref{sec:theory}), we obtain the theoretical predictions for the swimmer trajectory and the joint angle kinematics which are plotted in Fig. \ref{fig:SI_hor_swimming} for $12$ swimming cycles. As in the vertical swimming case there is very good agreement between the theoretical and experimental trajectories and also the average swimming speed (Fig. \ref{fig:SI_hor_swimming}b). There is good agreement also in the phase plot of the joint angles shown in Fig. \ref{fig:SI_hor_swimming}c. The deviation between theory and experiments for the tail-fork angle is possibly due to the fact that the organism is swimming in a curved trajectory (Fig. \ref{fig:SI_hor_swimming}a) which is seemingly due to the observed asymmetry in the tail-fork angle about the equilibrium value of $90 \degree$.
\subsection{Formulation of Theoretical Model for Swimming Cercariae}	
\label{sec:theory}
\begin{figure}
\begin{center}
\includegraphics[width=0.6\textwidth]{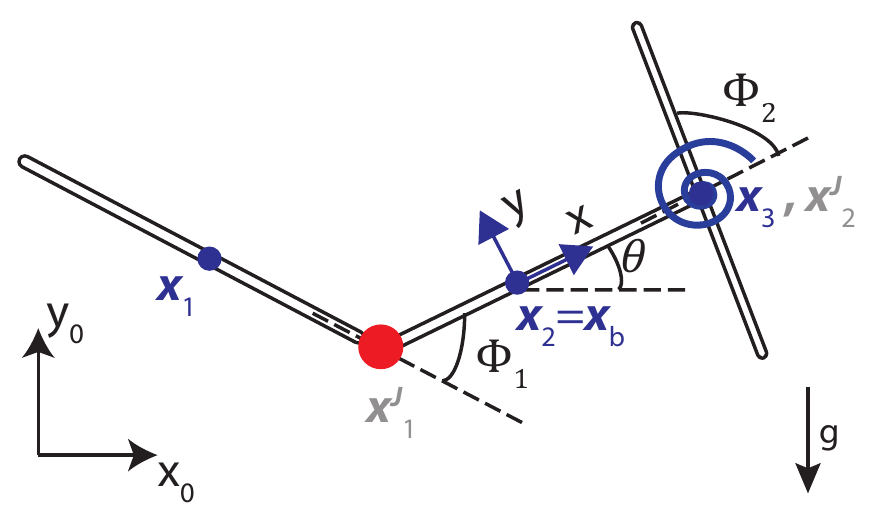} 
\caption{\label{fig:SI_figure_2} Schematic of the three-link swimmer ('T-swimmer') showing the coordinate system used in the formulating the model. The positions of the links (blue) and joints (gray) are shown. The joint between the longitudinal links is active (red dot), while the `T-joint' is a passive linear torsional spring (blue spiral). The direction of gravity is along the negative $y$ axis.
}
\end{center}
\end{figure}
 Here we formulate the three-link and four-link swimmer models described in the paper. The four-link model, being more representative of actual cercariae, is used to compare theoretical predictions with experiments. However, here we only formulate the model for the simpler three-link swimmer, the extension of which to a four-link-swimmer is straightforward.

 The model parameters for the three-link swimmer consists of the link lengths $l_1$, $l_2$ and $l_3$ and transverse dimensions $r_1$, $r_2$ and $r_3$; the amplitude of actuation of the active joint $A$ and the torsional stiffness of the passive joint $\Gamma$. The swimmer is immersed in an ambient fluid of viscosity $\mu$ and density $\rho_f$. Following from the low-Reynolds-number hydrodynamics that governs the swimming, the problem can be made dimensionless using the following scales: the length scale $l_c$ is taken as half the swimmer length (the swimmer length is defined as the distance from the tip of the head to the base of the fork when the tail is relaxed and straight); the time scale $t_c=1/f$ which is the inverse of the beat frequency; the typical force and torque scales are $\mu f l_c^2$ and $\mu f l_c^3$, respectively. Following normalization with the aforementioned scales, the system is governed by the dimensionless link lengths $\hat{l}_1$, $\hat{l}_2$ and $\hat{l}_3$; aspect ratios $\epsilon_1$, $\epsilon_2$ and $\epsilon_3$; and the actuation amplitude $A$. The only non-geometric parameter is the dimensionless torsional stiffness of the `T-joint' which is given by $\hat{\Gamma} = \Gamma / (\mu f l_c^3)$. 

 Following the approach of \cite{passov2012dynamics} (see also supplementary information), we write down the equations of motion governing the dynamical system consisting of the three-links. We specify a body-fixed reference frame whose origin is assumed to be the centre of link 2 (Fig. \ref{fig:SI_figure_2}), with the $x$ axis aligned with the longitudinal axis of the link, which is oriented at an angle $\theta$ with the lab-fixed frame of reference (Fig. \ref{fig:SI_figure_2}). The position and orientation of the body-fixed coordinate system is measured relative to a lab-fixed coordinate system whose origin is taken as $\boldsymbol{X_0}=\boldsymbol{0}$ (Fig. \ref{fig:SI_figure_2}). The positions of the links are given by the positions of their centers denoted by $\boldsymbol{x_i}$. The links are assumed rigid and the joints are assumed to have only a rotational degree of freedom. The angles between the links, at the two joints are denoted by $\phi_1$ and $\phi_2$. For the case of the three-link swimmer $\phi_1=\phi_t$ and $\phi_2=\phi_{tf}$, as defined in the main paper. Thus the position, orientation and shape of the swimmer is entirely defined by the state-vector $\{ \boldsymbol{X_b}, \boldsymbol{\Phi} \}$, where $\boldsymbol{X_b}=\{\boldsymbol{x_b}, \theta \}$ is the body position vector and $\boldsymbol{\Phi} = \{\phi_1, \phi_2\}$ shape vector. Since the swimmer is restricted to motions on a two-dimensional plane, the positions vectors are understood to mean $\boldsymbol{x} = \{ x, y , 0\}$. Therefore, the body-velocity of the swimmer is defined as $ \boldsymbol{V_b} = \{ \boldsymbol{v_b}, \omega_b \}$, where $\boldsymbol{v_b} = \boldsymbol{\dot{x}_b}$, $\omega_b = \dot{\theta}$ and the shape-velocity is given by $\boldsymbol{\dot{\Phi}} = \{\dot{\phi}_1, \dot{\phi}_2\}$, where the `dot' refers to a time derivative. Note that the velocity vectors, for instance $\boldsymbol{v_b}$ (denoted by bold symbols) correspond to elements in $\mathbb{R}^3$, unless specified otherwise. The angular velocity vectors have the form $\boldsymbol{\omega} = \omega \boldsymbol{\hat{e}_z}$, which is of the form of a scalar times a constant unit vector since the axis of rotation is always along $\boldsymbol{\hat{e}_z}$.

 Due to the rigid nature of the links and the continuity of translational velocities at the joints, we can write the translational and angular velocities of each link, denoted respectively by $\boldsymbol{\dot{x}_i}$ and $\boldsymbol{\omega_i}$, in terms of the translational velocity $\boldsymbol{V_b}$ of the body-fixed frame and shape velocity $\boldsymbol{\dot{\Phi}}$ of the joint angles:
 \begin{align}
\boldsymbol{\dot{x}_i} & = \boldsymbol{v_b} +  \omega_b \boldsymbol{\hat{e}_z} \wedge (\boldsymbol{x_i}-\boldsymbol{x_b}) + \sum_{j=1}^{2} \mathcal{L}_{ij} \dot{\phi}_{j} \boldsymbol{\hat{e}_z} \wedge (\boldsymbol{x_i}-\boldsymbol{x^{J}_j}) \label{eqn:trans_vel}\\
\boldsymbol{\omega_i}  & = \omega_b \boldsymbol{\hat{e}_z} + \sum_{j=1}^2 \mathcal{L}_{i j} \boldsymbol{\hat{e}_z} \dot{\phi}_{j} \label{eqn:ang_vel}
 \end{align}
where $\boldsymbol{\hat{e}_z}$ is the unit vector in the $z$ direction, $\boldsymbol{x^J}_j$ are the positions of the joints and $\mathcal{L}_{ij}$ is a matrix which depends on the topology of the swimmer and is $1$ if the velocity of $i^{th}$ link depends on rotation about the $j^{th}$ joint and is zero otherwise. For the case of a three-link swimmer, with two joints and link 2 as the body-frame this is given by $\mathcal{L} = \begin{bmatrix} 1  & 0 \\ 0  & 0 \\ 0 & 1 \end{bmatrix}$. $\wedge$ denotes the vector cross-product.

Equations (\ref{eqn:trans_vel}) and (\ref{eqn:ang_vel}) can be written in matrix form as:
 \begin{equation}
 	\boldsymbol{\dot{X}} = \mathcal{A} \boldsymbol{V_b} + \mathcal{B} \boldsymbol{\dot{\Phi}},
\end{equation}
where $\boldsymbol{\dot{X}} = [\boldsymbol{\dot{X_1}}, \boldsymbol{\dot{X_2}}, \boldsymbol{\dot{X_3}}]$ and $\boldsymbol{\dot{X_i}} = \{\boldsymbol{\dot{x}_i}, \omega_i\}$. $\mathcal{A}$ and $\mathcal{B}$, are, respectively, the matrices connecting the body-velocity and shape-velocity of the swimmer to the velocity of the $i^{th}$ link.

Once the kinematics relations between the links have been specified, we now specify the forces and torques on the links due to the fluid. Let $\boldsymbol{F_i}=\{\boldsymbol{f_i},  m_i \}$ and $\boldsymbol{F}= [\boldsymbol{F_1}, \boldsymbol{F_2}, \boldsymbol{F_3} ]$. From the linearity of the Stokes equations which govern the fluid flow around the swimmer at low-Reynolds-numbers, there exists a linear relationship between the velocities of the links and the resulting hydrodynamic forces \cite{lealbook} which gives:
\begin{equation}
\boldsymbol{F} = - \mathcal{R} \boldsymbol{\dot{X}}, \label{eqn:F}
\end{equation}
where $\mathcal{R}$ is the grand-resistance-matrix and in this work it is assumed to be block-diagonal, consisting of the resistance matrices $\boldsymbol{R}_i$ of each of the links. Such a block-diagonal structure arises since hydrodynamic interactions are neglected at leading order. Thus $\mathcal{R}$ is given by:
\begin{equation}
\mathcal{R}=
\begin{bmatrix}
\boldsymbol{R_1} & \boldsymbol{0} 	& \boldsymbol{0}\\
 \boldsymbol{0} & 	\boldsymbol{R_2}	& \boldsymbol{0} \\
 \boldsymbol{0} & \boldsymbol{0} & 	\boldsymbol{R_3}
\end{bmatrix}
\end{equation}
For slender bodies this is true to $O(1/ln(\epsilon))$, where $\epsilon = l/r$ is the aspect ratio (length to transverse dimension) of the slender body. Given the high amplitude beating of the tail in cercariae, this assumption is not strictly valid but provides a valuable leading order estimate. Extending the analysis to the next order in $1/ln(\epsilon)$ using slender-body-theory \cite{cox1970sbt} to capture hydrodynamic interactions between the links is straightforward and will be considered in a future work. 

Local slender-body-theory gives the net force on a slender rod of length $l$ and radius $r$ at a location $s$ along its axis as $\boldsymbol{f}(s) = -C_N \boldsymbol{v_n}(s) - C_L \boldsymbol{v_s}(s)$, where $\boldsymbol{v_n}(s)$ and $\boldsymbol{v_n}(s)$, denote the velocity in the direction normal and tangential to the axis of the slender body. $C_N$ and $C_L$ are the drag coefficients in the normal and tangential directions and for the limit of very slender bodies ($ \epsilon = l/r \gg 1$), they are given by $C_L= 2 \pi / ln(\epsilon)$ and $C_N = 4 \pi / ln(\epsilon)$. The torque on a slender rod due to its angular velocity $\omega$ is given by $\tau = -C_N \omega l^3 / 12 $. We can thus write the resistance matrix for a linear, rigid link as: 
\begin{equation}
\boldsymbol{R_i} = C_{L} l_i \begin{bmatrix}
\cos{\alpha_i}^2+\frac{C_N}{C_L} \sin{\alpha_i}^2 & -\sin{\alpha_i} \cos{\alpha_i}(\frac{C_N}{C_L}-1) & 0 \\
-\sin{\alpha_i} \cos{\alpha_i}(\frac{C_N}{C_L}-1) & \frac{C_N}{C_L} \cos{\alpha_i}^2+ \sin{\alpha_i}^2  & 0 \\
0 & 0 & \frac{C_N}{C_L}\frac{l_i^2}{12}
\end{bmatrix}
\end{equation}
Here $\alpha_i$ are the orientation angles of each link relative to the body-frame orientation. For the `T-swimmer', with link-2 as the body frame, $\alpha_1 =\theta -\phi_1$, $\alpha_2 =\theta$ and $\alpha_3 = \theta + \phi_2$ (see Fig. \ref{fig:SI_figure_2}). 

The net force on the swimmer and net moments at the location of the body-frame are given by $\boldsymbol{f_b} = \sum_{i=1}^{3} \boldsymbol{f_i}$ and $\boldsymbol{m_b} = \sum_{i=1}^{3} ( m_i \boldsymbol{\hat{e}_z} + (\boldsymbol{x_i}-\boldsymbol{x_b}) \wedge \boldsymbol{f_i})$, where $\boldsymbol{m_b} = m_b \boldsymbol{\hat{e}_z}$. In matrix form this is given by:
\begin{equation}
\boldsymbol{F_b} = 
\begin{bmatrix} \boldsymbol{f_b} \\
m_b
\end{bmatrix}
= \mathcal{A}^T \boldsymbol{F} 
\end{equation}
where the $^T$ denotes the matrix transpose. A consequence of Stokes flow is that the net hydrodynamic force and torque on a self-propelled swimmer are instantaneously zero. However, in our case, since we are trying model cercariae which are negatively buoyant, bottom-heavy swimmers, there exists an additional force and torque due to gravity. Taking the direction of gravity to be along the negative $y$ axis (see Fig. \ref{fig:SI_figure_2}), the net buoyant force and torque on the swimmer are $\boldsymbol{F}_{g} = [ \boldsymbol{f_g} , m_g],$ where $\boldsymbol{f_g} = \{0, (\rho_{cerc}- \rho_f) V_{cerc} g, 0\}$ is the resultant force due to gravitational and buoyant effects and $\boldsymbol{m_g} = m_g \boldsymbol{\hat{e}}_z =  \rho_f V_{cerc} g \boldsymbol{\hat{e}}_y \wedge (\boldsymbol{x}_{COG}-\boldsymbol{x}_{COB}) \sin{\theta}$ is the torque on the swimmer due to the centre-of-gravity and centre-of-buoyancy of the swimmer being separated (see Section \ref{sec:cog}). Using this expression for $\boldsymbol{F}_{g}$, the net force and torque balance for the swimmer gives:
\begin{align}
& \boldsymbol{F_b} +  \boldsymbol{F}_{g}  = \boldsymbol{0} \\
 & -\mathcal{A}^T \mathcal{R} (\mathcal{A} \boldsymbol{V_b} + \mathcal{B} \boldsymbol{\dot{\Phi}}) + \boldsymbol{F}_{g}  = \boldsymbol{0}
\end{align} 
Simplifying and inverting the relation to solve for $\boldsymbol{V_b}$, we have:
\begin{align}
   & \boldsymbol{V_b} = \mathcal{G}_{bb}^{-1}( -\mathcal{G}_{bs} \boldsymbol{\dot{\Phi}} + \boldsymbol{F}_{g} )
   \label{eqn:body_vel}
\end{align}
where $\mathcal{G}_{bb}= \mathcal{A}^T \mathcal{R} \mathcal{A}$ and $\mathcal{G}_{bs}= \mathcal{A}^T \mathcal{R} \mathcal{B}$. This expression thus gives the velocity of the body-frame attached to the swimmer at link-2 based on velocities of the joint angles. 

For the purposes of modelling a swimmer with a passive joint of fixed torsional stiffness, it is necessary to express the RHS of Equation (\ref{eqn:body_vel}) in terms of the internal torques at the joints. To do this, we consider a force and torque balance on a section of the swimmer which is divided at the $j^{th}$ joint. Thus the force and torque at the joint denoted by $\boldsymbol{f^{J}_j}$ and $\boldsymbol{m^{J}_j}$ are then balanced by the hydrodynamic forces and torques on the remaining portion of the swimmer. As per our convention earlier, the links that affect a given joint $j$ is given by $\mathcal{J}_j = \{i \mid \mathcal{L}_{ij}=1 \}$. The forces and torques at the joints can be written in terms of the bending moments and forces on the links as, $\boldsymbol{f^J}_j=-\sum \limits_{i \in \mathcal{J}_j} \boldsymbol{f_i}$ and $\boldsymbol{m^J}_{j} = -\sum \limits_{i \in \mathcal{J}_j} (\boldsymbol{m_i} + (\boldsymbol{x_i} - \boldsymbol{x^J_{j}}) \wedge \boldsymbol{f_i})$. Note that the effect of gravity does not enter into bending moments at the joints assuming each link is homogeneous in density, and hence has a coincident COG and COB (an assumption we make for cercariae as well (see Section \ref{sec:cog}). 

Let $\boldsymbol{\tau}=[\tau_1, \tau_2]$ be the vector containing the internal torques at the joints. It is straightforward to show that $\boldsymbol{\tau} = -\mathcal{B}^T \boldsymbol{F}$, where $\mathcal{B}$ is the same matrix appearing in Equation (\ref{eqn:body_vel}). Substituting for $\boldsymbol{F}$ from Equation (\ref{eqn:F}), we get:
\begin{equation}
\boldsymbol{\tau} = \mathcal{B}^T \mathcal{R} (\mathcal{A} \boldsymbol{V_b} + \mathcal{B} \boldsymbol{\dot{\Phi}})
\end{equation}
Substituting for $\boldsymbol{V_b}$ from Equation (\ref{eqn:body_vel}), we get:
\begin{equation}
\boldsymbol{\tau} = (\mathcal{G}_{ss} - \mathcal{G}^T_{bs} \mathcal{G}^{-1}_{bb} \mathcal{G}_{bs}) \boldsymbol{\dot{\Phi}}, \label{eqn:torque_angvel}
\end{equation}
where $\mathcal{G}_{ss} = \mathcal{B}^T \mathcal{R} \mathcal{B}$. To arrive at Equation (\ref{eqn:torque_angvel}) we have used $\mathcal{G}^T_{bs} \mathcal{G}^{-1}_{bb} \boldsymbol{F_g} = \boldsymbol{0}$ since gravitational buoyant forces do not contribute to bending moments at the joints as per our assumptions. Inverting this relation we obtain $\boldsymbol{\dot{\Phi}} = \mathcal{H} \boldsymbol{\tau}$, where $\mathcal{H} = (\mathcal{G}_{ss} - \mathcal{G}^T_{bs} \mathcal{G}^{-1}_{bb} \mathcal{G}_{bs})^{-1}$. We can now write the body velocity of the swimmer in terms of the torques at the joints as:
\begin{equation}
\boldsymbol{V_b} =  \mathcal{G}_{bb}^{-1}( -\mathcal{G}_{bs} \mathcal{H} \boldsymbol{\tau} + \boldsymbol{F}_{g}) \label{eqn:body_vel_torque}
\end{equation}
For our three-link swimmer, we specify the active joint to be undergoing a sinusoidal motion at a frequency $f$ and of the form $\phi_1 = A \sin{(2 \pi f t)}$. The passive joint, on the other hand, experiences a restoring torque due to the torsional spring when the angle deviates from the equilibrium angle, that is given by $\tau_2 = \Gamma (\phi_0 - \phi_2)$ where for the `T-swimmer' the equilibrium angle $\phi_0 = \pi/2$. We can calculate the torque required at the active link to produce the sinusoidal motion above by solving Equation (\ref{eqn:torque_angvel}) for $\tau_1$. Doing so we get:
\begin{equation}
\boldsymbol{\tau}=
\begin{bmatrix}
\frac{(2 \pi A \cos{(2 \pi f t)} - \mathcal{H}_{12} \Gamma (\pi/2 - \phi_2))}{\mathcal{H}_{11}} \\
\Gamma (\pi/2-\phi_2)
\end{bmatrix}
\label{eqn:tau}
\end{equation}

Equation (\ref{eqn:body_vel_torque}) is now solved with $\boldsymbol{\tau}$ as specified by Equation (\ref{eqn:tau}) by integrating forward in time using an adaptive Runge-Kutta solver.

An important parameter is the swimming efficiency which we define as: 
\begin{equation}
\eta_{dist} = \frac{X_{cycle}}{P_{diss,cycle}}, 
\end{equation}
 where $X_{cycle}$ is the displacement of the swimmer body-frame (link-2 in Fig. \ref{fig:SI_figure_2}) averaged over a swimming cycle and $P_{diss,cycle}$ is the net energy dissipated per cycle due to all the motions of the swimmer. This is given by $P_{diss,cycle} = - \int_{0}^{t_{cycle}}\boldsymbol{F} \cdot \boldsymbol{V} dt =  \int_{0}^{t_{cycle}} \boldsymbol{V}^T \mathcal{R}\boldsymbol{V} dt$, where $t_{cycle} = 1/f$ is the time to complete one swimming cycle.
\end{document}